\pdfoutput=1

\documentclass[aps,showpacs,nofootinbib,superscriptaddress]{revtex4}
\usepackage{graphicx}
\usepackage{subfigure}  
\usepackage{amsmath} 
\usepackage{amsfonts} 
\usepackage{amssymb} 
\usepackage{slashed} 
\usepackage{braket}
\usepackage{units}
\usepackage{color}
\allowdisplaybreaks

\newcommand{\sla}{\ensuremath{\slashed}}
\newcommand{\sle}{\ensuremath{\slashed\epsilon}}
\newcommand{\slp}{\ensuremath{\slashed p}}
\newcommand{\slq}{\ensuremath{\slashed q}}
\newcommand{\slz}{\ensuremath{\slashed z}}
\newcommand{\slk}{\ensuremath{\slashed k}}
\newcommand{\nn}{\nonumber}
\DeclareMathOperator\arctanh{arctanh}
\begin{document}
\title{$\Delta(1232)$ resonance in the $\vec{\gamma}p\rightarrow p\pi^0$ reaction at threshold}

\author{A. N. \surname{Hiller Blin}}
\affiliation{Departamento de F\'isica Te\'orica and IFIC, Centro Mixto
Universidad de Valencia-CSIC, Institutos de Investigaci\'on de
Paterna, E-46071 Valencia, Spain}
\author{T. \surname{Ledwig}}
\affiliation{Departamento de F\'isica Te\'orica and IFIC, Centro Mixto
Universidad de Valencia-CSIC, Institutos de Investigaci\'on de
Paterna, E-46071 Valencia, Spain}
\author{M. J.  \surname{Vicente Vacas}}
\affiliation{Departamento de F\'isica Te\'orica and IFIC, Centro Mixto
Universidad de Valencia-CSIC, Institutos de Investigaci\'on de
Paterna, E-46071 Valencia, Spain}

\today

\begin{abstract}
We calculate the neutral pion photoproduction on the proton near threshold in covariant baryon chiral perturbation theory, including the $\Delta(1232)$ resonance as an explicit degree of freedom, up to chiral order $p^{7/2}$ in the $\delta$ counting. We compare our results with recent low-energy data from the Mainz Microtron for angular distributions and photon asymmetries. The convergence of the chiral series of the covariant approach is found to improve substantially with the inclusion of the  $\Delta(1232)$ resonance.

\end{abstract}


\maketitle

\section{Introduction}\label{intro}

 Single pion photoproduction on nucleons  has been abundantly studied since the early fifties.
The first  low-energy theorems (LET) trying to describe the reaction close to threshold were obtained in the pioneering work of Kroll and Ruderman~\cite{Kroll} in a model-independent way by imposing gauge and Lorentz invariance.  Their results were later improved by using current algebra and 
the partial conservation of the axial-current~\cite{DeBaenst:1971hp,Vainshtein:1972ih}.  These LET were quite successful on the description of the charged pion channels but showed clear discrepancies with data for the  $\gamma + p \rightarrow p + \pi^0$ process~\cite{Mazzucato:1986dz,Beck:1990da, Drechsel:1992pn,Bernard:2006gx}. 
 
Bernard et al.~\cite{Bernard:1991rt,Bernard:1992nc} found some corrections coming from loop-diagram contributions in a Chiral Perturbation Theory (ChPT) calculation, which significantly reduced these discrepancies.  Later, they calculated the reaction in heavy-baryon ChPT (HBChPT). This approach,
introduced in Refs.~\cite{Jenkins:1990jv,Jenkins:1991es}, provides a systematic power-counting scheme solving the problems found in Ref.~\cite{Gasser:1987rb} for the loops with baryons.  For the data available at the time, the agreement with a fourth-order analysis was very good~\cite{Bernard:2001gz}.

However, the advent of new and high precision threshold data for both cross sections and photon asymmetries~\cite{Hornidge:2012ca}, from the Mainz Microtron (MAMI), showed that this approach is not sufficient for the full description of the process. In fact, the agreement of the HBChPT calculation up to $\mathcal{O}(p^4)$ with data is good only up to about $20~\unit{MeV}$ above threshold, as shown in Ref.~\cite{FernandezRamirez:2012nw}. For higher energies, the convergence is spoiled and would require an even higher-order calculation with many unknown Low Energy Constants (LECs). 

On the other hand, there are some alternative renormalization schemes to deal with the power-counting problem of the baryon loops. In particular, the Extended On Mass Shell (EOMS) ChPT~\cite{Gegelia:1999gf,Fuchs:2003qc}, though technically more complicated, keeps covariance and satisfies analyticity constraints, both lost in the HBChPT formulation. Furthermore, it usually converges faster.
 This model succeeded in describing processes like pion-scattering and many baryon observables in the low-energy regime. Examples are masses, magnetic moments, axial form factors, among others~\cite{Fuchs:2003ir,Lehnhart:2004vi,Schindler:2006it,Schindler:2006ha,Geng:2008mf,Geng:2009ik, MartinCamalich:2010fp,Alarcon:2011zs,
Ledwig:2012,Chen:2012nx,Ledwig:2013,Ledwig:2014rfa,Lensky:2014dda}. Unfortunately, as the description of the neutral pion photoproduction on protons goes, the fully covariant calculation up to fourth order~\cite{Hilt:2013uf} seems even slightly worse than what the HBChPT approaches had obtained so far.

A possibility to improve the convergence, which we explore in this work, is the explicit inclusion of the $\Delta(1232)$ resonance as an additional degree of freedom. At higher energies, the $\Delta$ clearly dominates the neutral pion photoproduction cross section~\cite{Ericson:1988gk}. Even close to threshold, its consideration could speed up the convergence of the chiral series if the size of the resonance tail is still large as compared to the purely nucleonic mechanisms. The possible relevance of the $\Delta$ mechanisms in our process was already suggested by Hemmert et al.~\cite{Hemmert:1996xg} and later in Refs.~\cite{Hornidge:2012ca,FernandezRamirez:2012nw}.
 
Recently,  the  $\Delta$ resonance has been included as a dynamic degree of freedom in many works. For instance,  Refs.~\cite{Pascalutsa:2004pk,Pascalutsa:2005vq,FernandezRamirez:2005iv} study pion electro- and photoproduction although their focus is at higher energies. There are also EOMS ChPT analysis of
Compton scattering~\cite{Lensky:2009uv,Blin:2015era} and $\pi N$ scattering~\cite{Alarcon:2012kn}. 
For the case of neutral pion photoproduction close to threshold,  the $\Delta$ mechanisms have been investigated 
in HBChPT~\cite{Hemmert:1996xg,Bernard:2001gz}, getting only moderate effects.  As was discussed in Ref.~\cite{Blin:2015}, this small effect could be due to the fact that both were static calculations, which omitted the fast energy dependence that comes from the full consideration of the $\Delta$  propagator. There is also a more recent work  in progress in  HBChPT at $\mathcal{O}(p^4)$ which shows a clear improvement when the $\Delta$ resonance is included~\cite{Cawthorne:2015}.

In Ref.~\cite{Blin:2015}, we studied the process $\gamma + p \rightarrow p + \pi^0$ in covariant ChPT, incorporating the $\Delta$ resonance as an explicit degree of freedom.  The calculation was of chiral order $p^3$ in the $\delta$ counting, which will be discussed below. This amounts to a nucleonic sector with tree-level and loop diagrams, but only tree diagrams containing  $\Delta$. Furthermore, the $\Delta$ pieces only depended on  two relatively well known couplings, $g_M$ and $h_A$. The results were very promising and showed a good agreement with data for both differential cross sections and asymmetries up to above 200 MeV.

 In this work, we extend the calculation to the next order in the $\delta$ counting, namely $\mathcal{O}(p^{7/2})$, 
which basically adds loop diagrams with $\Delta$ propagators. The loop diagrams do not require any additional coupling. There is only one new LEC, $g_E$, which appears in a tree diagram and is poorly known. Furthermore, we are able to describe the process consistently with LECs
which are mostly constrained by other observables. 

The paper is organized as follows:
In Sec.~\ref{Samps}, we present the basic formalism to extract the neutral pion production channel's amplitudes and observables. In Sec.~\ref{theory}, we introduce all the theoretical tools necessary for our calculation. This includes the ChPT Lagrangians, as well as the renormalization and power-counting scheme used. In Sec.~\ref{results}, we show and discuss our results for cross sections, photon asymmetries and multipoles. Finally, the summary and outlook are given in Sec.~\ref{summary}.

\section{Basic formalism}\label{Samps}

\begin{center}
\begin{figure}
\includegraphics[width=.2\textwidth]{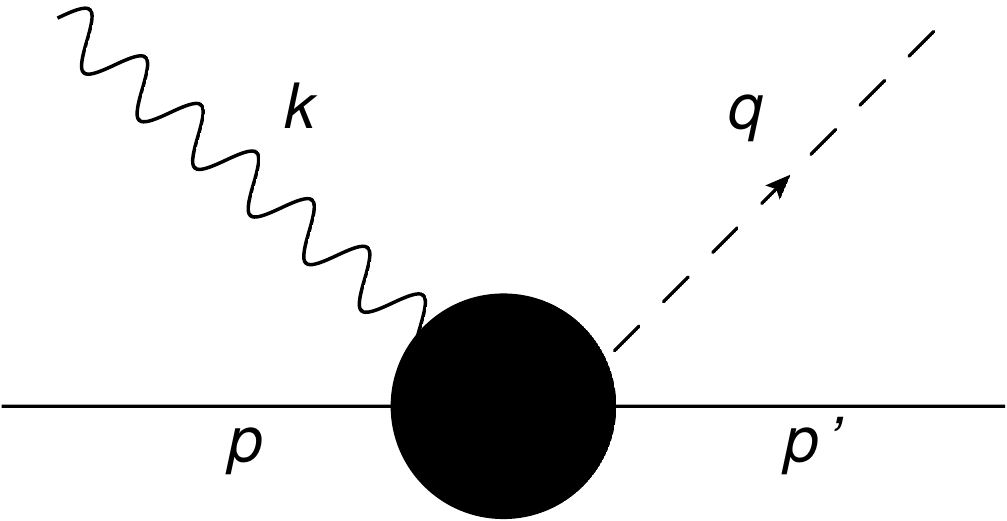}
\caption{Generic representation of the pion photoproduction process. The incoming photon and proton momenta are given by $k$ and $p$, while those of the outgoing neutral pion and proton are denoted by $q$ and $p'$, respectively.}
\label{fBlobProc}
\end{figure}
\end{center}

The process we are studying is represented in Fig.~\ref{fBlobProc}. The four-momenta $k$, $q$, $p$ and $p'$ belong to the photon, $\pi^0$, incoming and outgoing protons, respectively.

We parameterize the scattering amplitude $\mathcal{M}$ as
\begin{align}
\epsilon_\mu \mathcal{M}^\mu =&\bar u(p')\left(V_Nq\cdot\epsilon\gamma_5+V_Kq\cdot\epsilon\slashed{k}\gamma_5+V_E\slashed{\epsilon}\gamma_5+V_{EK}\slashed{\epsilon}\slashed{k}\gamma_5\right)u(p),
\end{align}
where $V_N$, $V_K$, $V_E$ and $V_{EK}$ are complex structure functions of the photon energy $k_\gamma$ in the laboratory frame and the angle $\theta$ between incoming photon and outgoing pion. The Dirac spinors $u(p)$ and $\bar u (p') = u^\dagger(p')\gamma_0$ are those of the nucleon in the initial and final states, respectively, and $\epsilon$ is the photon polarization.
%
%

Another commonly used representation is explicitly current conserving by definition and has the form
\begin{align}
\nn \epsilon_\mu \mathcal{M}^\mu = &\epsilon_\mu \bar u(p')\left(
	\sum_{i=1}^4A_iM_i^\mu\right)u(p),
\intertext{with}
\nn \epsilon\cdot M_1 =& \mathrm{i} \slk\sle\gamma_5,\\
 \nn \epsilon\cdot M_2 =&\mathrm{i}(p'\cdot\epsilon k\cdot q - q\cdot\epsilon k\cdot(p+p'))\gamma_5,\\
\nn \epsilon\cdot M_3 =&\mathrm{i}(\sle k\cdot q- \slk q\cdot\epsilon)\gamma_5,\\
 \nn \epsilon\cdot M_4 =&\mathrm{i}(\sle k \cdot(p+p')-\slk p'\cdot \epsilon - 2m \slk\sle)\gamma_5.
\intertext{Here, $m$ is the nucleon mass. Note that in the center-of-mass system $p\cdot\epsilon=0$. The conversion between parameterizations is straightforward:}
\nn A_1 = & 
\mathrm{i}\left(V_{EK}-\frac{m}{k\cdot p}\left(V_E+k\cdot q V_K\right)\right),\\
\nn A_2 = &
\mathrm{i}\frac{V_N}{2 k\cdot p},\\
\nn A_3 = &
\mathrm{i}\left(V_K\left(1-\frac{k\cdot q}{2k\cdot p}\right)-\frac{V_E}{2k\cdot p}\right),\\
\nn A_4 = &
-\frac{\mathrm{i}}{2k\cdot p}\left(V_E+k\cdot qV_K\right).
\end{align}

Finally, for the calculation of multipoles it is convenient to use the equivalent representation in terms of the Chew--Goldberger--Low--Nambu (CGLM) amplitudes~\cite{Chew:1957tf}, 
\begin{align}
\nn \epsilon_\mu \mathcal{M}^\mu = \frac{4\pi W}{m}\chi_f^\dagger\mathcal{F}\chi_i,
\end{align}
where $\chi_i$ and $\chi_f$ are the initial and final state Pauli spinors, respectively, and $W=\sqrt{s}$ is the center-of-mass energy. For real photons and in the Coulomb gauge, the amplitude $\mathcal{F}$ may be written as
\begin{align}
\nn \mathcal{F} = \mathrm{i}\vec\sigma\cdot\vec\epsilon\mathcal{F}_1
+  \vec\sigma\cdot\hat q\vec\sigma\cdot\hat k\times\vec\epsilon\mathcal{F}_2
+  \mathrm{i}\vec\sigma\cdot\hat k \hat q\cdot \epsilon\mathcal{F}_3
+  \mathrm{i}\vec\sigma\cdot\hat q \hat q\cdot \epsilon\mathcal{F}_4,
\end{align}
with $\vec\sigma$ the Pauli matrices. The conversion between parameterizations is given by
\begin{align}
\nn& \epsilon_\mu \bar u(p')\left(
	\sum_{i=1}^4A_iM_i^\mu\right)u(p) = \frac{4\pi W}{m} \chi_f^\dagger\mathcal{F}\chi_i,
\intertext{with}
\nn 
\mathcal{F}_1 =&\frac{\sqrt{(E_i+m)(E_f+m)}}{8\pi W}\Bigg[
-\left(k_0 +\frac{k_0^2}{E_i+m}\right)A_1-k\cdot q A_3\\
\nn&+\left(-k_0^2+2k_0m+\frac{2k_0^2m}{E_i+m}-k_0(E_i+E_f)-k_0|\vec q|\cos\theta\right)A_4\Bigg],\\
\nn\\
 \nn\mathcal{F}_2 =&\frac{\sqrt{(E_i+m)(E_f+m)}}{8\pi W}|\vec q|\Bigg[\left(\frac{k_0}{E_f+m} +\frac{k_0^2}{(E_i+m)(E_f+m)}\right)A_1\\
\nn&
-\frac{k_0k\cdot q}{(E_i+m)(E_f+m)}A_3\\
\nn&-\left(k_0\frac{k_0^2+2k_0m+k_0(E_i+E_f)+k_0|\vec q|\cos\theta}{(E_i+m)(E_f+m)}
+\frac{2k_0m}{E_f+m}\right)A_4\Bigg],\\
\nn\\
\nn\mathcal{F}_3 =& \frac{\sqrt{(E_i+m)(E_f+m)}}{8\pi W}|\vec q|\Bigg[
-k_0^2\frac{E_i+E_f + k_0 + q_0}{E_i+m}A_2\\
\nn&+\left(k_0+\frac{k_0^2}{E_i+m}\right)  (A_4 - A_3)
\Bigg],
\\\nn\\
\nn\mathcal{F}_4 =&\frac{\sqrt{(E_i+m)(E_f+m)}}{8\pi W}|\vec q|^2\Bigg[
\left(k_0\frac{k_0+E_i+E_f+q_0}{E_f+m}\right)A_2\\
\nn &+\left(\frac{k_0}{E_f+m}+\frac{k_0^2}{(E_i+m)(E_f+m)}\right) (A_4 - A_3)
\Bigg].
\end{align}

We compare our model with the full set of data  of 
Refs.~\cite{Hornidge:2012ca,Hilt:2013uf} on the unpolarized angular cross section
\begin{equation}
\frac{\mathrm{d}\sigma}{\mathrm{d}\Omega}=\frac{|\vec{q}|m^2}{2\pi W(s-m^2)}\sum_\epsilon{
\frac{\text{Tr}\left[
\mathcal{M}^*\cdot(\slashed{p}'+m)\cdot\mathcal{M}\cdot(\slashed{p}+m)
\right]}
{2}},
\end{equation}
 and $\Sigma$, the linearly polarized photon asymmetry
\begin{equation}
\Sigma=\frac{\mathrm{d}\sigma_\perp-\mathrm{d}\sigma_\parallel}{\mathrm{d}\sigma_\perp+\mathrm{d}\sigma_\parallel},
\end{equation}
with $d\sigma_\perp$ and $d\sigma_\parallel$ the angular cross sections for photon polarizations perpendicular and parallel to the reaction plane, respectively. 

In the CGLM representation, the differential cross section and photon asymmetry are usually written with the help of the response functions
\begin{align}
\nn R_T =& |\mathcal{F}_1|^2 + |\mathcal{F}_2|^2 + \frac12\sin^2\theta\left(|\mathcal{F}_3|^2+|\mathcal{F}_4|^2\right)
\\\nn&- \text{Re}\left[
2\cos\theta \mathcal{F}_1^*\mathcal{F}2- \sin^2\theta\left(
\mathcal{F}_1^*\mathcal{F}_4+\mathcal{F}_2^*\mathcal{F}_3+\cos\theta \mathcal{F}_3^*\mathcal{F}_4\right)\right]
\intertext{and}
\nn R_{TT} =& \frac12\sin^2\theta\left(|\mathcal{F}_3|^2+|\mathcal{F}_4|^2\right)
\\\nn&+ \text{Re}\left[ \sin^2\theta\left(
\mathcal{F}_1^*\mathcal{F}_4+\mathcal{F}_2^*\mathcal{F}_3+\cos\theta \mathcal{F}_3^*\mathcal{F}_4\right)\right],
\end{align}
with which one obtains
\begin{align}
\nn\frac{\mathrm{d}\sigma}{\mathrm{d}\Omega_\pi}=\frac{|\vec q|}{k_\gamma}R_T
\hspace{5mm}\text{and}\hspace{5mm}
\Sigma=-\frac{R_{TT}}{R_T}.
\end{align}
The lowest multipoles $E_{0+}$, $M_{1+}$, $M_{1-}$ and $E_{1+}$ read~\cite{Bernard:1992nc}:
\begingroup
\renewcommand*{\arraystretch}{1.5}{
\begin{align*}
\left(
\begin{array}{c}
E_{0+}\\
M_{1+}\\
M_{1-}\\
E_{1+}
\end{array}\right)
=\int_{-1}^1\mathrm{d}x
\left(\begin{array}{cccc}
\frac{1}{2}P_0(x) & -\frac{1}{2}P_1(x) & 0 & \frac{1}{6}\left[P_0(x)-P_2(x)\right]\\
\frac{1}{4}P_1(x) & -\frac{1}{4}P_2(x) & \frac{1}{12}\left[P_2(x)-P_0(x)\right] & 0\\
-\frac{1}{2}P_1(x) & \frac{1}{2}P_0(x) & \frac{1}{6}\left[P_0(x)-P_2(x)\right] & 0\\
\frac{1}{4}P_1(x) & -\frac{1}{4}P_2(x) & \frac{1}{12}\left[P_0(x)-P_2(x)\right] & \frac{1}{10}\left[P_1(x)-P_3(x)\right]
\end{array}\right)
\left(
\begin{array}{c}
\mathcal{F}_1(x)\\
\mathcal{F}_2(x)\\
\mathcal{F}_3(x)\\
\mathcal{F}_4(x)
\end{array}\right),
\end{align*}
}\endgroup
where $x=\cos(\theta)$ and $P_l$ are the Legendre polynomials. We furthermore use the reduced multipoles
\begin{align}
\bar{M}_{1\pm}=\frac{M_{1\pm}}{|\vec{q}|}
\hspace{5mm}\text{and}\hspace{5mm}
\bar{E}_{1+}=\frac{E_{1+}}{|\vec{q}|},
\end{align}
as for energies close to threshold these multipoles are linearly related to the absolute value of the pion momentum.

\section{Theoretical model}\label{theory}

We will analyse the  MAMI pion photoproduction data~\cite{Hornidge:2012ca,Hilt:2013uf} using a fully covariant ChPT framework and including the $\Delta(1232)$ resonance as an explicit degree of freedom. While the baryon ChPT power-counting problem~\cite{Gasser:1987rb} is solved in the EOMS scheme, additional special care is needed when taking 
this spin-$3/2$ resonance into account. Besides the pion mass and the external momenta, another small parameter appears, $\delta= M_\Delta - m\approx 300~\unit{MeV}$, which is heavier than $m_\pi\sim140~\unit{MeV}$, but small when compared to the spontaneous symmetry-breaking scale $\Lambda\sim m$. In the low-energy range of our study, we count $\delta^2$ as being of $\mathcal{O}(p)$, following Ref.~\cite{Lensky:2009uv}. Thus one obtains the power-counting rule
\begin{align}
D=4L+\sum_{k=1}^{\infty}{kV^{k}}-2N_\pi-N_N-\frac12N_\Delta,
\end{align}
according to which is given the order $D$ of a diagram with $L$ loops, $V^{k}$ vertices from $\mathcal{L}^{(k)}$, $N_\pi$ pionic propagators, $N_N$ nucleonic propagators and $N_\Delta$ $\Delta(1232)$ propagators. In Ref.~\cite{Blin:2015}, we presented a calculation up to order $p^3$. A tree diagram of order $p^{7/2}$ and proportional to $g_E$ was also investigated.
 Our aim here is to extend the model up to order $p^{7/2}$. That only amounts to the consideration of new loop diagrams with $\Delta$ propagators. Thus, no further low-energy constants  are required.

 We start with the relevant terms of the Lagrangian for the neutral pion production on the proton with real photons, including only pions, nucleons and photons as degrees of freedom. We follow the naming conventions for the LECs introduced in Ref.~\cite{Fettes:2000gb}. At first order we have  
\begin{equation}
\mathcal{L}_N^{(1)}=\bar{\Psi}\left(\mathrm{i}\sla{\mathrm{D}}-m+\frac{g_0}{2}\sla u\gamma_5\right)\Psi,
\end{equation}
where $\Psi$ is the nucleon doublet $\left(p,n\right)$ with mass $m$ and $\mathrm{D}_\mu=\left(\partial_\mu+\Gamma_\mu\right)$ is the covariant derivative with
\begin{equation*}
\Gamma_\mu=\frac12\left[u^\dagger(\partial_\mu-\mathrm{i}r_\mu)u
+u(\partial_\mu-\mathrm{i}l_\mu)u^\dagger\right]. 
\end{equation*}
At a $\mathcal{O}(p)$ calculation, the low-energy constant $g_0$ corresponds to the axial-vector coupling constant $g_A=1.27$. The meson fields appear through 
\begin{equation*}
u=\exp\left(\frac{\mathrm{i}\phi}{2F}\right), \quad
 \phi=\left(\begin{array}{cc}
\pi^0&\sqrt2\pi^+\\
\sqrt2\pi^-&-\pi^0
\end{array}\right),
\end{equation*}
where at $\mathcal{O}(p)$ $F$ corresponds to the pion decay constant $F_\pi$ with numerical value $92.4~\unit{MeV}$, and also in 
$
u_\mu=\mathrm{i}\left[u^\dagger(\partial_\mu-\mathrm{i}r_\mu)u
-u(\partial_\mu-\mathrm{i}l_\mu)u^\dagger\right].
$
 The photon field $\mathcal{A}_\mu$ couples through 
\begin{equation*}
r_\mu=l_\mu=\frac e2\mathcal{A}_\mu(\mathbb{I}_2+\tau_3),
\end{equation*}
where $\tau_3$ is the Pauli matrix and $e$ is the (negative) electron charge. At second order, the only relevant terms are 
\begin{equation}
\mathcal{L}_N^{(2)}=\frac{1}{8m}\bar{\Psi}\left(
c_6 f_{\mu\nu}^+ + c_7\text{Tr}\left[f_{\mu\nu}^+\right]
\right)\sigma^{\mu\nu}\Psi+\dots,
\end{equation}
where $ f_{\mu\nu}^+=uf_{\mu\nu}^Lu^\dagger+u^\dagger f_{\mu\nu}^Ru$ and for our case
$
f_{\mu\nu}^{R}=f_{\mu\nu}^{L}=\partial_\mu r_\nu - \partial_\nu r_\mu -\mathrm{i}\left[r_\mu,r_\nu\right].
$
The tensor $\sigma^{\mu\nu}$ is given by
$
\frac{\mathrm{i}}{2}\left[\gamma^\mu,\gamma^\nu\right]
$. In the particular case of the $\gamma+p\rightarrow p+\pi^0$ scattering amplitude, the LECs $c_6$ and $c_7$ appear only as a combination $\tilde{c}_{67}=c_6+c_7$. This constant can be fixed from the nucleons' magnetic moments. Using the model of Ref.~\cite{Ledwig:2012}  leads to the value $\tilde{c}_{67}=2.3$ at $\mathcal{O}(p^3)$ and $\tilde{c}_{67}=2.5$ when $\Delta$ loops are included\footnote{In Ref.~\cite{Ledwig:2012},  only the isovector combination was presented.}.
Finally, at third order we have 
\begin{align}
 \mathcal{L}_N^{(3)}= &d_8\frac{\mathrm{i}}{2m}\left\{
\bar\Psi\varepsilon^{\mu\nu\alpha\beta}\text{Tr}\left[\tilde f_{\mu\nu}^+u_\alpha\right]
\mathrm{D}_\beta\Psi
\right\}+\text{h.c.}\\
 +&d_9\frac{\mathrm{i}}{2m}\left\{
\bar\Psi\varepsilon^{\mu\nu\alpha\beta}\text{Tr}\left[f_{\mu\nu}^+
\right]u_\alpha
\mathrm{D}_\beta\Psi
\right\}+\text{h.c.}\nonumber\\
 +&d_{16}\frac{1}{2}\left\{
\bar\Psi\gamma^{\mu}\gamma_5\text{Tr}[\chi_+]u_\mu\Psi
\right\}\nonumber\\
 +&d_{18}\frac{\mathrm{i}}{2}\left\{
\bar\Psi\gamma^{\mu}\gamma_5[\mathrm{D}_\mu,\chi_-]\Psi
\right\}+\dots,\nonumber
\end{align}
where
$
 \tilde f_{\mu\nu}^+ = f_{\mu\nu}^+-\frac12\text{Tr}[f_{\mu\nu}^+]$ and  
$
\chi_\pm=u^\dagger\chi u^\dagger\pm u\chi^\dagger u$. We will work in the isospin limit as was done in Ref.~\cite{Hilt:2013uf}, hence taking $\chi=m_\pi^2$, the pion mass squared\footnote{The corrections to the approximation of using a single pion mass and also a single nucleon mass for the loop calculations is of higher order. Nevertheless, doing so we cannot study the cusp effects appearing at the opening of the charged pion channels.}. We use the convention $\varepsilon^{0123}=-\varepsilon_{0123}=-1$. Here, the LECs appear in the combinations $\tilde{d}_{89}=d_8+d_9$ and $\tilde{d}_{168}=2d_{16}-d_{18}$.

ChPT was initially developed for interactions between mesons and photons~\cite{Weinberg:1978kz,Gasser:1983yg,Gasser:1984gg}. The leading-order Lagrangian for this kind of interactions is given by
\begin{align}
\mathcal{L}_\pi^{(2)}=\frac{F^2}{4}\text{Tr}{\left[
\mathrm{D}_\mu U(\mathrm{D}^\mu U)^\dagger+\chi U^\dagger +U \chi^\dagger
\right]},
\end{align}
where $U=u^2$ and whose covariant derivative acts as
$
\mathrm{D}_\mu U=\partial_\mu U - \mathrm{i}r_\mu U + \mathrm{i}U l_\mu.
$

To describe the $\Delta$ interactions  we use consistent Lagrangians which ensure the decoupling of the spurious spin-$1/2$ components of the Rarita-Schwinger field $\Delta=\left(\Delta^{++},\Delta^{+},\Delta^{0},\Delta^{-}\right)$~\cite{Pascalutsa:1998pw,Pascalutsa:1999zz,Pascalutsa:2000kd,Pascalutsa:2006up}. The relevant pieces are
\begin{align}
\mathcal{L}^{(1)}_{\Delta\pi N}=&\frac{\mathrm{i}h_A}{2FM_\Delta}\bar{\Psi}T^a\gamma^{\mu\nu\lambda}(\partial_\mu\Delta_\nu)(\mathrm{D}_\lambda^{ab}\pi^a) + \text{H.c.},\\
  \mathcal{L}^{(2)}_{\Delta\pi N}=&\frac{h_1}{2FM_\Delta^2}\bar{\Psi}T^a\gamma^{\mu\nu\lambda}(\partial_\lambda\slashed\partial\pi^a)(\partial_\mu\Delta_\nu) + \text{H.c.},\\
  \mathcal{L}^{(2)}_{\Delta\gamma N}=&\frac{3\mathrm{i}eg_M}{2m(m+M_\Delta)}\bar{\Psi}T^3(\partial_\mu\Delta_\nu)\tilde{F}^ {\mu\nu} + \text{H.c.},\\
  \mathcal{L}^{(3)}_{\Delta\gamma N}=&-\frac{3eg_E}{2m(m+M_\Delta)}\bar{\Psi}T^3\gamma_5(\partial_\mu\Delta_\nu)F^ {\mu\nu} + \text{H.c.},
\end{align}
where the tensor $\gamma^{\mu\nu\lambda}$ reads $\frac14\left\{\left[\gamma^\mu,\gamma^\nu\right],\gamma^\lambda\right\}$ and  the covariant derivative 
$\mathrm{D}_\lambda^{ab}\pi^a=\delta^{ab}\partial_\lambda\pi^b - \mathrm{i}eQ_\pi^{ab}\mathcal{A}_\lambda\pi^b$, with $Q_\pi^{ab}=-\mathrm{i}\epsilon^{ab3}$. The electromagnetic field and its dual are given by $F^{\mu\nu}=\partial^{\mu}\mathcal{A}^{\nu}-\partial^{\nu}\mathcal{A}^{\mu}$ and $\tilde{F}^{\mu\nu} = \frac12\epsilon^{\mu\nu\alpha\beta}F_{\alpha\beta}$, respectively. There are two couplings for the pion ($h_A$, $h_1$) and two for the photon, the magnetic piece ($g_M$) of chiral order two and the electric piece ($g_E$) of order three. At third order, the Lagrangian contains an additional $\gamma N\Delta $ Coulomb coupling  which vanishes for real photons. As the value for $h_1$ has been found to be consistent with zero~\cite{Pascalutsa:2006up}, we neglect this piece in our calculation. The value for $h_A$ can be directly obtained from the $\Delta$ width, while $g_M$ and $g_E$ were obtained fitting  pion electromagnetic production at energies around the resonance peak. The conventions and definitions for the isospin operators $T$ follow Ref.~\cite{Pascalutsa:2006up}:
\begin{align*}
T^1=\frac{1}{\sqrt 6}\left(
\begin{array}{cccc}
-\sqrt 3 &0&1&0
\\
0&-1&0&\sqrt 3
\end{array}
\right),
\hspace{5mm}
T^2=\frac{-\mathrm{i}}{\sqrt 6}\left(
\begin{array}{cccc}
\sqrt 3 &0&1&0
\\
0&1&0&\sqrt 3
\end{array}
\right),
\hspace{5mm}
T^3=\sqrt{\frac23 }\left(
\begin{array}{cccc}
0 &1&0&0
\\
0&0&1&0
\end{array}
\right).
\end{align*}

\begin{figure}
\subfigure[]{
\label{fO1a}
\includegraphics[width=0.3\textwidth]{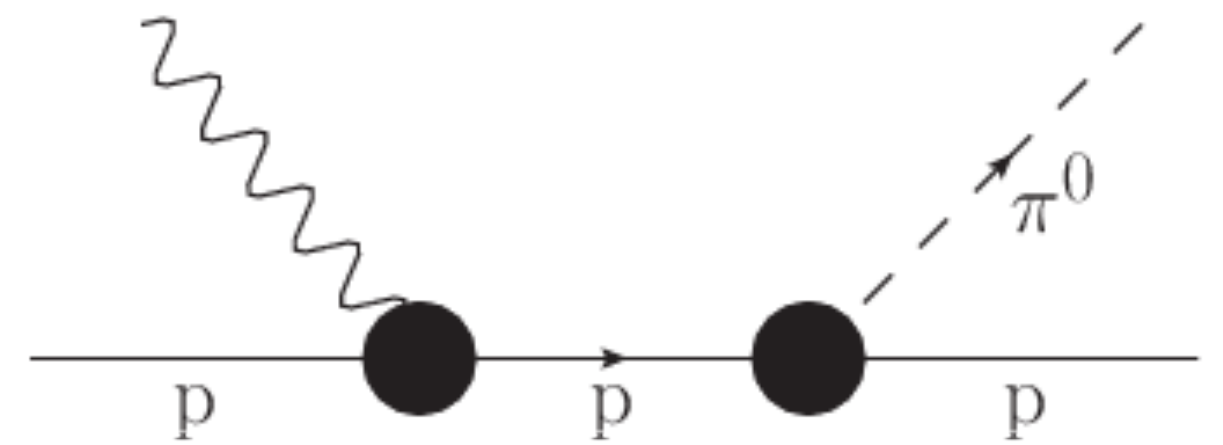}}
\subfigure[]{
\label{fO1c}
\includegraphics[width=0.2\textwidth]{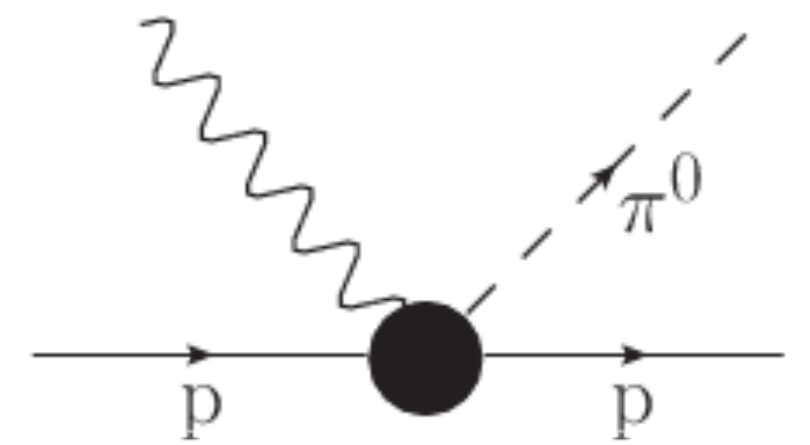}}
\caption{Tree diagrams for  $\pi^0$ photoproduction off protons. Crossed terms are also included in the calculation. The black dots represent vertices of chiral order 1 to 3. Diagram b) starts at order 3.}
\label{FTree12}
\end{figure}

\begin{figure}[htb]
\begin{center}
\subfigure[]{
\label{fDd}
\includegraphics[width=0.45\textwidth]{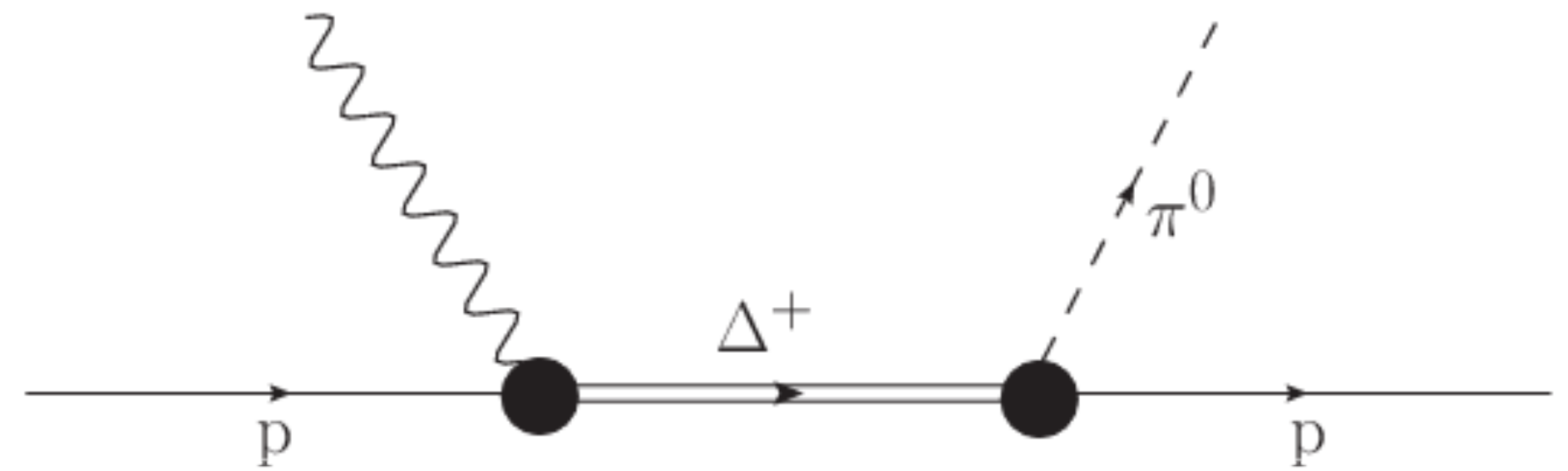}}
\end{center}
\caption{$\Delta$ tree diagram for $\pi^0$ photoproduction off protons. The crossed term is also included in the calculation.}\label{FTree32}
\end{figure}

In Figs.~\ref{FTree12} and~\ref{FTree32}, we show the tree-level diagrams. 
The full set of loop diagrams contributing to the considered channel up to $\mathcal{O}(p^{7/2})$ can be found depicted in Figs.~\ref{FLoop12} and~\ref{FLoop32} in App.~\ref{ADiagrams}. They have been evaluated applying the EOMS renormalization scheme, with the help of FORM~\cite{FORM:2000,FORM:2012} and FeynCalc~\cite{FC:1991,FC:2016}. First, 
we have removed the infinities using the modified minimal subtraction $(\widetilde{MS})$ scheme \cite{Scherer:2012xha}. Then, after making an expansion of the amplitudes\footnote{The chosen expansion parameters were, as in Ref.~\cite{Alarcon:2012kn}, $m_\pi$, $\nu=(s-u)/(4 m)$ with $s$ and $u$ the Mandelstam variables of $\mathcal{O}(p)$, and the Mandelstam variable $t$ of order $\mathcal{O}(p^2)$, as well as the parameter $\delta$ explained above.},
we have also absorbed the power-counting breaking terms into LECs. Obviously, those diagrams from Fig.~\ref{FLoop12} which exclusively contain mesonic loops do not break the power counting. The analytical expression obtained for the power-counting breaking terms in the nucleonic sector reads
\begin{align*}
\frac{\mathrm{i} e g_A^3 m}{32F_\pi^3 \pi^2}\left[
\left(4\nu - 3\frac{m_\pi^2}{\nu}\right)\sle\gamma_5
+\left(3 - 3\frac{m_\pi^2}{\nu^2}\right)\sle\slk\gamma_5
+ \frac{1}{\nu}q\cdot\epsilon\slk\gamma_5
-\frac{2m}{\nu}q\cdot\epsilon\gamma_5
\right].
\end{align*}
The additional power-counting breaking terms coming from the introduction of the $\Delta$ loops are obtained analogously, but have large expressions which are therefore not shown here.

\begin{figure}
\begin{center}
\includegraphics[width=0.3\textwidth]{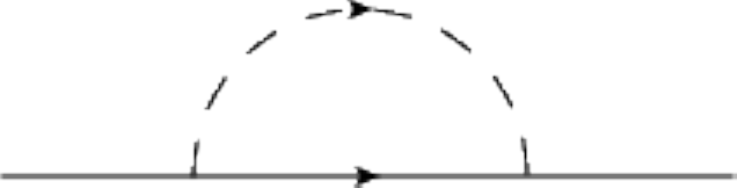}\hspace{1.5cm}
\includegraphics[width=0.3\textwidth]{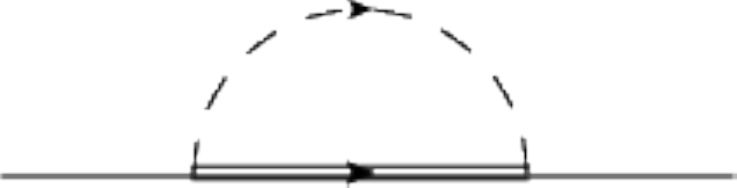}
\end{center}
\caption{Diagrams contributing to the proton's self-energy.}
\label{FSelfEn}
\end{figure}

In order to systematically take into account all the higher-order contributions up to the studied order $\mathcal{O}(p^{7/2})$, the wave-function renormalization (WFR)  was taken into account for the external proton and pion legs of the tree diagrams of $\mathcal{O}(p^1)$, as the correction amounts to multiplying this tree-level amplitude by $Z_p\sqrt{Z_\pi}$, which adds corrections of $\mathcal{O}(p^2)$. All the corrections to higher-order amplitudes or to the external photon leg would be at least of $\mathcal{O}(p^4)$. The analytical expression for this correction factor when including only nucleonic intermediate states reads 
\begin{align}\label{wfr}
    Z_p = \frac{1}{1-\Sigma_p'}\Big|_{\slashed p = m}=1-\frac{3g_A^2m_\pi^2}{32\pi^2F_\pi^2m^2(4m^2-m_\pi^2)}\Bigg[
2m_\pi(m_\pi^2-3m^2){\sqrt{4m^2-m_\pi^2}}\arccos\left(\frac{m_\pi}{2m}\right)\nonumber\\
+(m_\pi^2-4m^2)\left((2m_\pi^2-3m^2)\log\left(\frac{m_\pi}{m}\right)-2m^2\right)
\Bigg]+\mathcal{O}(p^4),
\end{align}
where $\Sigma_p$ is the self energy of the proton. Since we are considering the $\Delta(1232)$ as an intermediate state, we also have to take into account this additional self-energy loop that enters the wave-function renormalization. Also in this case, we took the $\mathcal{O}(p^2)$ term and added it to Eq.~\ref{wfr}. The analytical expression for this piece $Z_p^\Delta$ can be found in Appendix~\ref{AWfrDel}. The self-energy diagrams for the proton external legs are depicted in Fig.~\ref{FSelfEn}.

As for the pion-leg WFR and renormalization of the pion-decay constant,
 we use the well-known expansions from Ref.~\cite{Gasser:1983yg}:
\begin{align}
\nn F_\pi &= F+\frac{m_\pi^2}{F}\left[L_4-\frac{1}{16\pi^2}\log\left(\frac{m_\pi^2}{m^2}\right)\right]+\mathcal{O}(p^3),\\
Z_\pi &= 1-\frac{m_\pi^2}{F_\pi^2}\left[2L_4+\frac{1}{16\pi^2}\log\left(\frac{m_\pi^2}{m_N^2}\right)\right]+\mathcal{O}(p^3).
\end{align}
Then, for the $\mathcal{O}(p)$ diagrams, the appearing factor $\sqrt{Z_\pi}/F$ can be expanded around the pion mass:
\begin{align}
\frac{\sqrt{Z_\pi}}{F}=\frac{1}{F_\pi}-\frac{3m_\pi^2\log\left(\frac{m_\pi^2}{m^2}\right)}{32\pi^2F_\pi^3}+\mathcal{O}(p^3),
\end{align}
therefore leading to an expression which up to the considered order does not depend on $L_4$ anymore.

Also the other low-energy constants appearing in the leading-order Lagrangian have to be corrected up to the considered order. This means that the nucleon mass $m$ in the nucleon propagator of the leading-order tree-level diagrams has to be calculated with corrections coming from higher-order self-energy loops. The contributions to the physical nucleon mass coming from the loops in Fig.~\ref{FSelfEn} are given by
\begin{align}
m_N = m - 4c_1 m_\pi^2-\frac{3g_A^2m_\pi^3}{64\pi^2F_\pi^2}\left[\frac{m_\pi}{m}\log\left(\frac{m_\pi^2}{m^2}\right)-4\sqrt{1-\frac{m_\pi^2}{4m^2}}\arccos\left(\frac{m_\pi}{2m}\right)\right]+m_N^\Delta+\mathcal{O}(p^4),\label{eq:16}
\end{align}
where $m_N^\Delta$ is the correction arising from the loop with a $\Delta$ propagator. Its expression can be found in Appendix~\ref{AWfrDel}.
The $\mathcal{O}(p^2)$ correction to $m_N$ is consequently approximately given by
\begin{align}
m_2=m-4c_1m_\pi^2 = m_N+\frac{3g_A^2m_\pi^3}{64\pi^2F_\pi^2}\left[\frac{m_\pi}{m_N}\log\left(\frac{m_\pi^2}{m_N^2}\right)-4\sqrt{1-\frac{m_\pi^2}{4m_N^2}}\arccos\left(\frac{m_\pi}{2m_N}\right)\right]-m_N^\Delta+\mathcal{O}(p^4).\label{eq:17}
\end{align}

Finally, the EOMS-renormalized expression for $g_A$, when including nucleonic intermediate states only, is given by~\cite{Schindler:2006it,Alarcon:2012kn}
\begin{align}
\nn g_A = g_0 + 4m_\pi^2d_{16}-\frac{g_Am_\pi^2}{16\pi^2F_\pi^2m^2}\Bigg[
\frac{(3g_A^2+2)m_\pi^3-8(g_A^2+1)m^2m_\pi}{\sqrt{4m^2-m_\pi^2}}\arccos\left(\frac{m_\pi}{2m}\right)\\
+(3g_A^2+2)m^2+((4g_A^2+2)m^2-(3g_A^2+2)m_\pi^2)\log\left(\frac{m_\pi}{m}\right)
\Bigg]+\mathcal{O}(p^{7/2}).\label{eqGarenorm}
\end{align}
The inclusion of the $\Delta(1232)$-loop diagrams leads to further corrections to $g_A$. They have been analyzed in an EOMS $SU(3)$ calculation~\cite{Ledwig:2014rfa}, leading to small contributions (of the order of 5 to 10 \%). Here, we have not considered these corrections, which in our case would mean just a shift of the parameter $d_{18}$ without otherwise affecting the quality of the fit\footnote{In our calculation,  Eq.~\ref{eqGarenorm} is just used to determine $d_{16}$ from $g_A$ and the fit parameter $g_0$. The constant $d_{16}$ only enters in the evaluation of two tree diagrams of $\mathcal{O}(p^3)$, always in combination with $d_{18}$.}.

We opted to consistently introduce the corrections to the constants in the Lagrangians by applying them only to the first-order tree-level diagrams: There we took $g_0$ for the axial-vector coupling, $m_2$ for the propagator mass, $F$ for the pion-decay constant, and we multiplied the wave-function renormalization. For all the higher-order tree and loop diagrams, we took the physical constants $g_A$, $m_N$ and $F_\pi$, as otherwise we would be introducing corrections of order higher than $\mathcal{O}(p^{7/2})$. Furthermore, this scheme allows for a better comparison with the results obtained in the EOMS $\mathcal{O}(p^3)$  calculation of pion-nucleon scattering of Alarcon et al.~\cite{Alarcon:2012kn}.  The analytical expressions of the amplitude can be found in App.~\ref{ADiagrams}.

\section{Results and discussion}\label{results}

We compare the theoretical model introduced in the previous sections to the experimental data from Ref.~\cite{Hornidge:2012ca}. Data points were taken for the linearly polarized photon asymmetry and differential cross section for an energy range from pion-production threshold up to over 200 MeV with an unprecedented precision. We will first review the $\mathcal{O}(p^3)$ calculation that was already studied in Ref.~\cite{Blin:2015}. In that work, the aim was to establish the relevance of the $\Delta$ degree of freedom for the neutral pion photoproduction, even close to threshold. Here, we will pay more attention to the consistency of our results (LECs) with other calculations using the EOMS scheme and of the same chiral order.

\subsection{$\mathcal{O}(p^3)$}

At this order, only  tree diagrams and the loop diagrams from Fig.~\ref{FLoop12} contribute. The loop diagrams from  Fig.~\ref{FLoop32}, which include $\Delta$ propagators, start at  $\mathcal{O}(p^{7/2})$. Also, the mass and coupling-constant corrections of Eqs.~\ref{eq:16}-\ref{eqGarenorm} can be truncated at order $p^3$. As previously discussed, we fix $g_A$, $m_N$ and $F_\pi$ to their physical values and use them for all except the lowest-order diagrams. The  $\pi N \Delta$ coupling  $h_A$, which is basically determined by the $\Delta$ width, was fixed to  $2.85$~\cite{Pascalutsa:2005vq}. 
The  $\gamma N \Delta$ coupling $g_E$, which leads to an $\mathcal{O}(p^{7/2})$ contribution, has been set to zero. The constant  $g_0$ has been fixed to the value obtained at the same chiral order in Ref.~\cite{Ledwig:2014rfa} ($g_0=1.16$). The remaining LECs $\tilde{c}_{67}$, $\tilde{d}_{89}$, $\tilde{d}_{168}$ and $g_M$ are left as fitting parameters. 

 Table~\ref{tab:0} shows the results of the fit at this order. As in Ref.~\cite{Blin:2015}, the agreement with data is excellent and the $\chi$-squared value is very low. The parameter $\tilde{c}_{67}$ converges to the value required by the nucleon magnetic moment ($\tilde{c}_{67}=2.3$~\cite{Ledwig:2012} as discussed in the previous section).
The LEC $\tilde{d}_{89}$, for which we don't have any alternative estimation in the particular renormalization scheme we use here, gets a value of natural size. Finally, in our calculation the LECs $d_{16}$ and $d_{18}$ always appear in the combination $\tilde{d}_{168}=2d_{16}-d_{18}$ and the individual constants cannot be disentangled. Actually, in the amplitude, at tree level, they are also fully correlated with $g_0$. This can be clearly seen by studying the error correlation matrix in fits that include the $g_0$ as a free variable.  As an example, fixing $g_0=1.05$, quite a reasonable value~\cite{Ledwig:2014rfa}, modifies  $m_N^2\,\tilde{d}_{168}$ from -10.1 to -6.9, while maintaining the other LECs and producing the same $\chi^2$. Using Eq.~\ref{eqGarenorm} and setting $g_A$ to its physical value, we can estimate  $d_{16}$ and thus calculate $d_{18}$. This would lead to  positive values for $d_{18}$ in disagreement with other calculations~\cite{Alarcon:2012kn}. However, we have checked that this particular result is very sensitive to choices, like the use of $g_A$ vs. $g_0$ for loops or the application of the wave-function renormalization for the higher-order diagrams, even when these choices amount to $\mathcal{O}(p^4)$ corrections. 
We have also estimated the size of the effects of the $\mathcal{O}(p^4)$ contributions by including the contact terms of that order in the amplitude. The expressions can be obtained from Appendix C of Ref.~\cite{Hilt:2013fda}. 
We have found that $\tilde{d}_{168}$ is very sensitive to the $\tilde{e}_{48}$, $\tilde{e}_{50}$ or $\tilde{e}_{112}$ LECs. For instance, taking $\tilde{e}_{48}=-4.5~\unit{GeV^{-3}}$ modifies $m_N^2\cdot\tilde{d}_{168}$ from $-10.1$ to $-0.4$ and leads to $d_{18}$ values negative and consistent with other works~\cite{Alarcon:2012kn}. The other constants and the $\chi^2$ are barely affected.

 Finally, 
the $\Delta$ coupling $g_M$ obtained in the fit is consistent  with the value given in Ref.~\cite{Pascalutsa:2005vq} in a study of pion electroproduction in the $\Delta$ region, as well as with the value of $g_M=3.16\pm0.16$ 
obtained from the $\Delta$ electromagnetic decay in Ref.~\cite{Blin:2015era}. We understand this as meaning that the neutral pion photoproduction data are sensitive to the size of the $\Delta$ contribution even at threshold. We have also checked that when including the  $g_E$ piece, of $\mathcal{O}(p^{7/2})$ and present in Ref.~\cite{Pascalutsa:2005vq}, the fit result for $g_M$ changes to 2.9 and $g_E=-1$ in full agreement with the aforementioned work. 

 We also tried to do a fit without the inclusion of the $\Delta(1232)$ diagrams. We were able to confirm the results shown in Ref.~\cite{Hilt:2013uf}. Namely, in an EOMS calculation it is impossible to reproduce the experimental steep growth of the differential cross section with the photon  energy at this order. The inclusion of the $\Delta(1232)$ degrees of freedom strongly improves the agreement with data up to energies higher than $200~\unit{MeV}$.

\begin{table}
\begin{center}
\begin{tabular}{|c|c|c|c|c|c|}
\hline 
  $g_0$ &$\tilde{c}_{67}$& $\tilde{d}_{89}\cdot m_N^2$ & $\tilde{d}_{168}\cdot  m_N^2$& $g_M$& $\chi^2$/d.o.f. \\ 
\hline 
 {\bf 1.16} & 2.32  & 1.28 & -10.1 & 3.08 & 0.79 \\
\hline 
\end{tabular} 
\end{center}
\caption{LEC values for the $\mathcal{O}(p^3)$ calculation. Fixed values are in boldface.}\label{tab:0}
\end{table}

\subsection{Full model at $\mathcal{O}(p^{7/2})$}

Next, we have added all contributions of $\mathcal{O}(p^{7/2})$ in the $\delta$-counting.
This amounts  to  the $\Delta$ tree diagram with the $g_E$ coupling, and the loop diagrams with $\Delta(1232)$ of Fig.~\ref{FLoop32}.
 All these loop amplitudes depend only on LECs that already appear at $\mathcal{O}(p^3)$. Thus,  $g_E$ is the only new additional LEC. We already explored its role in Ref.~\cite{Blin:2015} and found that its contribution was small. 

As in the previous section, the value for the constant $g_0$ has been taken from Ref.~\cite{Ledwig:2014rfa}. In its model with the $\Delta$ resonance, $g_0$ varies between $1.05$ and $1.08$.
The remaining LECs, $\tilde{c}_{67}$,  $\tilde{d}_{89}$,  $\tilde{d}_{168}$, $g_M$ and $g_E$ are left as fitting parameters. The results of the fit are shown in Table~\ref{tab:1}.
\begin{table}
\begin{center}
\begin{tabular}{|c|c|c|c|c|c|c|}
\hline 
   $g_0$&$\tilde{c}_{67}$& $\tilde{d}_{89}\cdot m_N^2$ & $\tilde{d}_{168}\cdot  m_N^2$& 
$g_M$& $g_E$& $\chi^2$/d.o.f. \\ 
\hline 
 1.05  & 2.45 & 1.67 & -9.7 & 2.28 & 3.30& 0.80 \\
 1.05  & 2.29 & 1.17 & -10.4 &{\bf 2.90} &  3.53 & 0.96 \\
\hline 
\end{tabular} 
\end{center}
\caption{LEC values in different versions of the $\mathcal{O}(p^{7/2})$ model. Fixed values appear in boldface.}\label{tab:1}
\end{table}
The first observation is that the quality of the fit is similar to the lower-order calculation. This happens even though we have an additional
 LEC. Therefore, the contribution of the new loop terms does not improve the agreement with data. This is reflected in the $g_M$ parameter, which affects the $\Delta$ mechanisms, that goes towards lower values. Also the $\pi N \Delta$ coupling $h_A$ prefers smaller values and the $\chi^2$ would sensibly decrease if we allowed for a 10$\%$ reduction of this constant. However, we prefer to keep the well established result obtained from the $\Delta$  width. The values found in the literature for $g_M$, using the same Lagrangian as in the present work, vary from $2.6\pm0.2$~\cite{Pascalutsa:2002pi} in a heavy-baryon calculation of Compton scattering to $g_M=2.8\pm0.2$~\cite{Pascalutsa:2004pk} (pion photoproduction), $g_M=2.9$~\cite{Pascalutsa:2005ts,Pascalutsa:2005vq} (pion electroproduction) and $g_M=3.16\pm0.16$~\cite{Blin:2015era} ($\Delta$ electromagnetic decay). The latter two, which correspond to covariant chiral calculations, prefer the larger values. We obtain a relatively low result, but we find that  fixing $g_M=2.9$ the quality of the fit would still be reasonable. 
This behavior is very consistent with power counting, as obtaining $g_M$ from the $\Delta$
electromagnetic decay amounts to a leading-order approximation, which is sufficient for the $\mathcal{O}(p^3)$ calculation of
the previous section. However, next-to-leading order effects in the determination of $g_M$ would also enter
in this $\mathcal{O}(p^{7/2})$ calculation of the pion photoproduction and a reasonable deviation 
from $g_M=3.16\pm 0.16$ could be expected.
%
The parameter $g_E$ is less well known, and numbers ranging from 2 to -7 can be found~\cite{Pascalutsa:2002pi,Pascalutsa:2004pk,Pascalutsa:2005vq}, although the later works prefer $g_E=-1$. Oppositely to the lower-order calculation~\cite{Blin:2015}, this term is  relevant and its absence  worsens the fit. Without the $\mathcal{O}(p^{7/2})$ loop mechanisms our fit also converges to $g_E=-1$ as stated above. However, when these higher-order terms are included, $g_E$ prefers positive values. This term is relevant for the $E_1^+$ multipole and we have found that the contribution, close to threshold, of the loop terms is very important in our model.

The changes from the $\mathcal{O}(p^3)$ calculation have been rather mild for the parameters $\tilde{d}_{89}$ and $\tilde{c}_{67}$. In particular, it is interesting that  $\tilde{c}_{67}$ is slightly larger. This change and the final value are consistent with the results of Ref.~\cite{Ledwig:2012} when the $\Delta$ loops were included.  Additionally, these LECs do not have strong correlations with the other parameters of the fit. 


However, the $\tilde{d}_{168}$ parameter is strongly correlated to $g_0$.
 Changes of the order of 10$\%$ in  $g_0$ lead to changes of  30$\%$ in $\tilde{d}_{168}$ without modifying neither the $\chi^2$ nor the other constants' values.  
As already discussed in the  previous section, $\tilde{d}_{168}$ is also very sensitive to higher-order contributions.  To estimate their effects, we have included some of the  $\mathcal{O}(p^{4})$ contact terms in our fit. For instance, the consideration of the term porportional to $\tilde{e}_{48}$, when choosing $\tilde{e}_{48}=-6.0~\unit{GeV^{-3}}$, leads to $m_N^2\cdot\tilde{d}_{168}=3.1$ (and thus a negative $d_{18} $),  $m_N^2\cdot\tilde{d}_{89}=1.1$, $g_M=2.9$, $g_E=2.1$ and $\chi^2/\text{d.o.f}=0.67$. Namely, most LECs are quite stable except for $g_E$ that changes by 40 \% and $\tilde{d}_{168}$ that is strongly modified. Similar results are obtained including the other contact terms. Therefore, we should expect large changes for these two parameters in a higher order calculation.

\begin{figure}[h]
\begin{center}
\includegraphics[width=\textwidth]{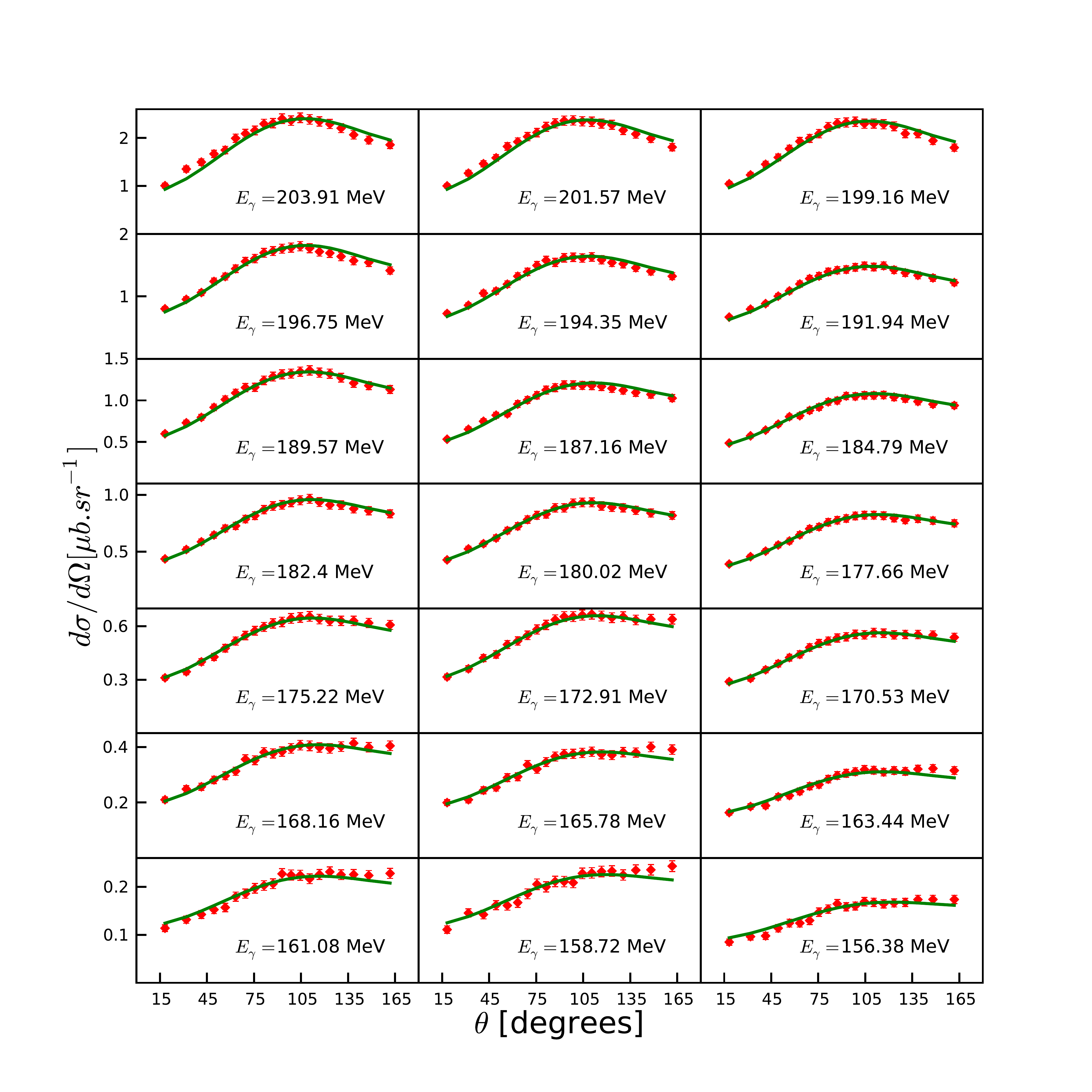}
\end{center}
\caption{Differential cross section as a function of the pion angle at different energies. Solid line: Best-fit theoretical model at $\mathcal{O}(p^{7/2})$. Experimental points from Refs.~\cite{Hornidge:2012ca, Hornidge:pri}.}\label{fig:11}
\end{figure}
 In Fig.~\ref{fig:11}, we show the angular differential cross section of our best fit at $\mathcal{O}(p^{7/2})$
versus the experimental data. Notice the quite small error bars of the data and the overall good agreement with the model for cross sections that vary more than one order of magnitude. The distributions are basically backward peaked. At the higher energies, there is a slight but systematic under-/overestimation at forward and backward angles respectively.

\begin{figure}[h]
\begin{center}
\includegraphics[width=\textwidth]{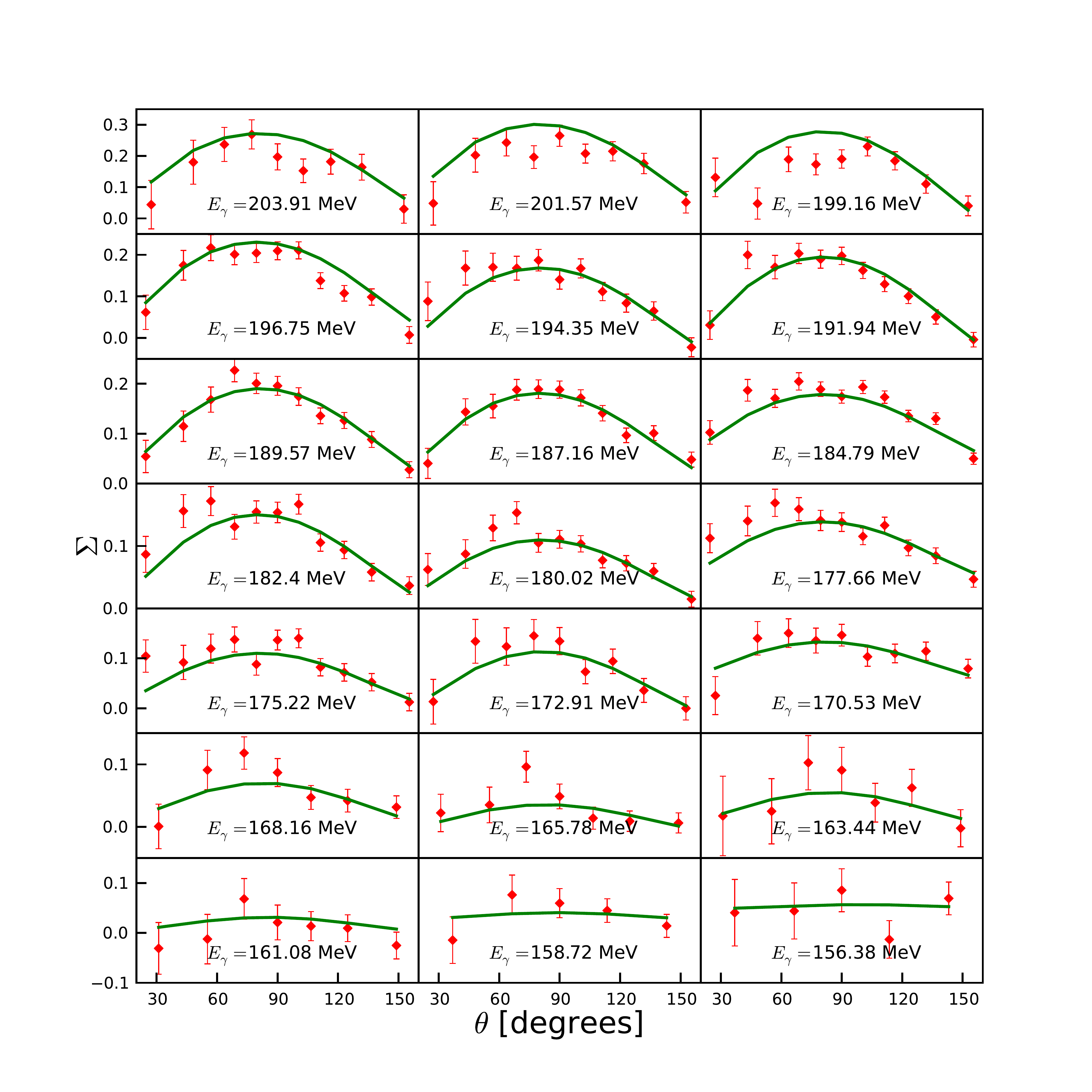}
\end{center}
\caption{Photon asymmetry as a function of the pion angle at different energies. Solid line: Best-fit theoretical model at $\mathcal{O}(p^{7/2})$. Experimental points from Refs.~\cite{Hornidge:2012ca, Hornidge:pri}.}\label{fig:12}
\end{figure}

 The linear photon asymmetries have been plotted in Fig.~\ref{fig:12}. Although the experimental uncertainties are larger, they also provide a very stringent test on the models, especially as the signal grows as a function of the photon energy. In fact, even though the number of data is much smaller and the error bars are larger  than for the angular distributions, its contribution to the full $\chi^2$ is similar. This may reflect the quality of data but could also point out some shortcoming of the model.

\begin{figure}[h]
\begin{center}
\includegraphics[width=\textwidth]{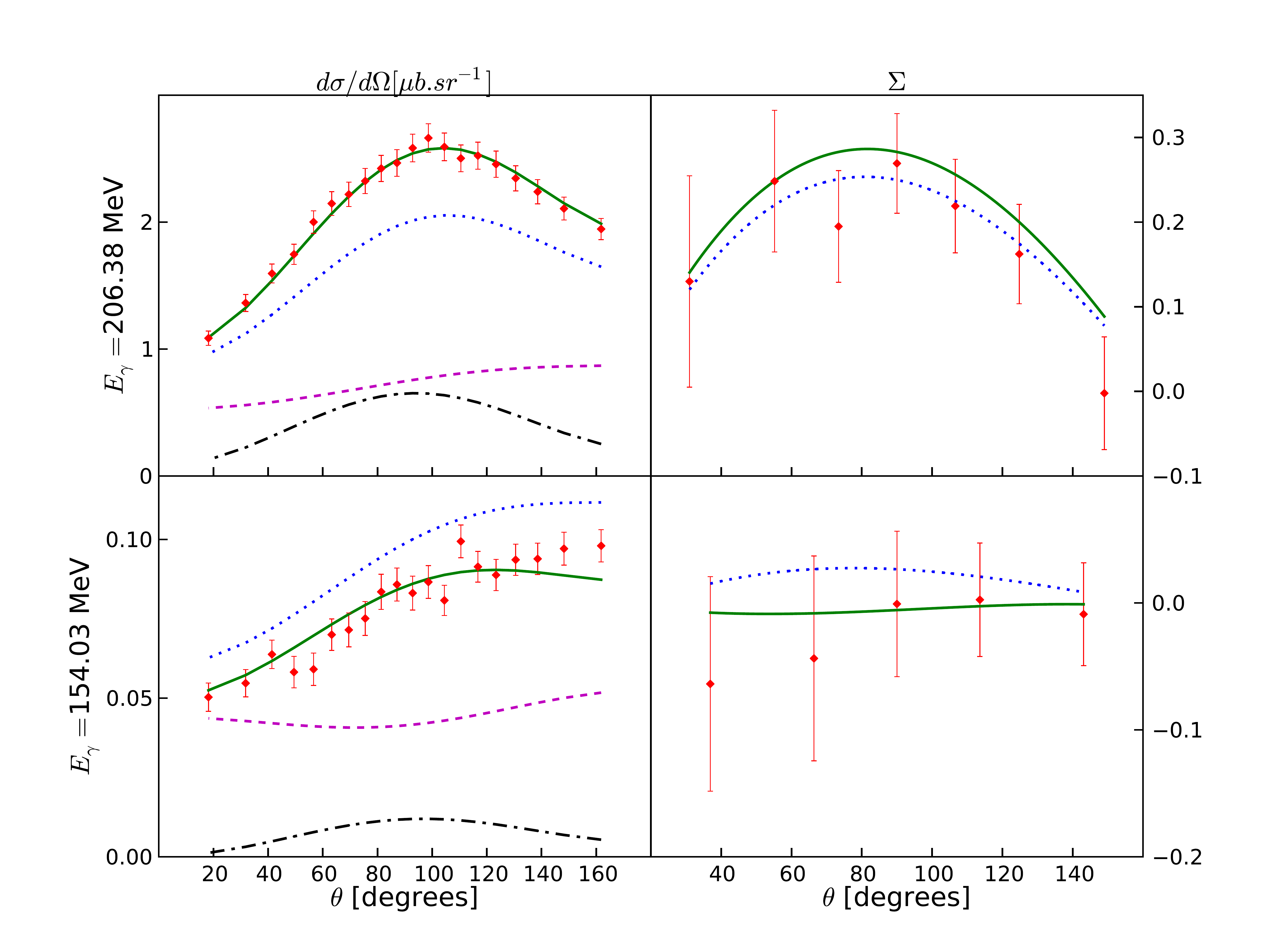}
\end{center}
\caption{Photon asymmetry and differential cross section as a function of the pion angle at two different energies: close to threshold and at above $200~\unit{MeV}$. Solid line: full model; dashed line: full model without $\Delta$; dash-dotted line: only $\Delta$;
dotted line: best nucleonic fit (without $\Delta$). Experimental points from Refs.~\cite{Hornidge:2012ca, Hornidge:pri}.}\label{fig:13}
\end{figure}

In Fig.~\ref{fig:13}, the calculation at two extreme energies of the data set is depicted. Here, we plot the results  of the full model without $\Delta$ and of the $\Delta$ diagrams alone (always with the parameter set of our best fit). A fit of $\mathcal{O}(p^3)$ including only nucleonic mechanisms is also shown. 

  Let's first discuss the purely nucleonic fit. As mentioned before, it is impossible to get a good fit at $\mathcal{O}(p^3)$ within our model. The reason is clear from the figure, whereas the asymmetry and the shape of the angular distribution are acceptably reproduced, the energy dependence is not strong enough, and the fit overestimates the low-energy data and underestimates the high energy ones. A higher-order calculation is mandatory for this $\Delta$-less approach.

 As soon as the $\Delta$ is incorporated the situation radically changes.  The relative size  of the $\Delta$ mechanisms is much larger at high energies and this helps to reproduce the energy dependence of the cross section. The detailed shape, and size, depends on the interference of the two kinds of mechanisms. 

\subsection{Multipoles}

\begin{figure}[h]
\begin{center}
\includegraphics[width=0.9\textwidth]{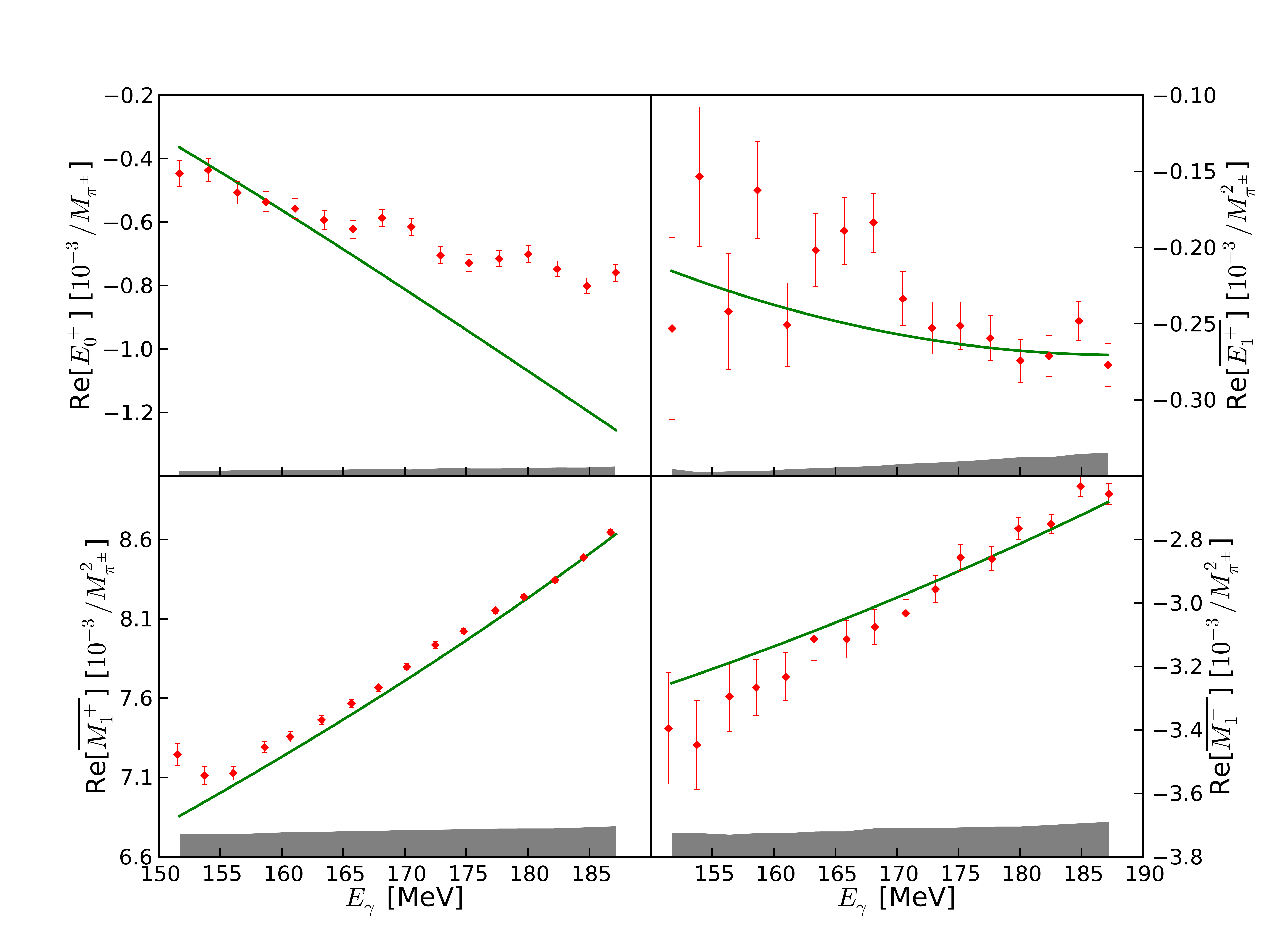}
\end{center}
\caption{Real part of S and P-wave multipoles from Ref.~\cite{Hornidge:2012ca} --- see also Ref.~\cite{FernandezRamirez:2012nw} --- vs. our full-model  calculation, as a function of the photon energy.
The error bars are only statistical errors.
 The gray band above the energy axis shows the systematic error of data~\cite{Hornidge:2012ca}.
}\label{fig:17}
\end{figure}

In Fig.~\ref{fig:17}, we compare our model with the empirical multipoles from Ref.~\cite{Hornidge:2012ca}. There were some assumptions in the extraction of their values. The imaginary parts of the $P$-wave multipoles were neglected, which is consistent with what we obtain in our model. For the imaginary part of $E_0^+$, which can be fixed from unitarity, it was found that it leads to smaller uncertainties than the statistical errors.
Another important source of uncertainty, mainly for $E_0^+$, is the influence of $D$-waves that can be sizable and grows fast as one departs from threshold. 

The calculated $E_1^+$, $M_1^+$ and $M_1^-$  multipoles agree well with the empirical ones. The quality of the agreement  for $M_1^-$ and  $E_1^+$ is similar to that of the $\mathcal{O}(p^4)$ covariant ChPT calculation from Ref.~\cite{Hilt:2013uf}. However, we reproduce well the large $M_1^+$ multipole 
which gets a substantially lower slope in Ref.~\cite{Hilt:2013uf}, a higher-order calculation. This fact can be explained by the absence in their model of the explicit inclusion of the $\Delta$, which plays a major role in this multipole.

For the $E_0^+$ case, we reproduce well the empirical estimation close to threshold, but our model leads to higher absolute values at larger energies.  A similar trend is observed in the $\mathcal{O}(p^4)$ purely nucleonic calculation~\cite{Hilt:2013uf}, although the discrepancy is lower in this case.
 As mentioned before, there is some uncertainty in the extraction of this multipole due to the presence of $D$-waves. The relevance of this partial wave has been explored in Refs.~\cite{FernandezRamirez:2009su,FernandezRamirez:2009jb}. They found that its contributions could seriously compromise the analysis and extraction of $E_0^+$, see e.g. Fig.~3 from Ref.~\cite{FernandezRamirez:2009su}. In our calculation, we have a relatively small $D$-wave contribution coming fundamentally from the crossed tree diagrams. As discussed in the appendix C of Ref.~\cite{Hilt:2013uf}, there could be large $D$-wave contributions coming from the $\mathcal{O}(p^4)$ Lagrangian terms. They could strongly modify $E^-_2$ that mixes with $E_0^+$, and  the changes could be large enough as to solve the discrepancies at large energies.

\section{Summary}\label{summary}

We have studied the neutral pion photoproduction on the proton at low energies in covariant ChPT with the explicit inclusion of the $\Delta(1232)$ resonance. We have used the EOMS renormalization scheme and made a full calculation up to  order $p^{7/2}$ in the $\delta$ counting. Comparing the $\mathcal{O}(p^{7/2})$ and  $\mathcal{O}(p^{3})$ calculations, we have found a good chiral convergence, in the sense that changes are quite small. However, as pointed out in previous works, even at the low energies discussed here, some $\mathcal{O}(p^4)$ contributions could be relevant.
 For instance, in our model, there is a quite small $D$-wave. The consideration of higher-order terms could modify that and, indirectly, affect the extraction of the $E^+_0$ multipole.

The model agrees well with the differential cross-section and photon-asymmetry data of Ref.~\cite{Hornidge:2012ca}, from threshold up to above 200 MeV. This extends the range of convergence from previous works of a higher chiral order, $\mathcal{O}(p^{4})$, in both HB and covariant ChPT. Our model without $\Delta$ only reproduces data very close to threshold, confirming the results from Ref.~\cite{Hilt:2013uf}, and showing that the improvement is basically due to the consideration of the $\Delta(1232)$ mechanisms.

 This is a nontrivial outcome of our work, because the LECs are mostly constrained by other observables. In particular,  $g_0$, $\tilde{c}_{67}$, $h_A$ and $g_M$  are bound by the nucleon axial-vector coupling, the proton magnetic moment, the strong and the elecromagnetic decays of the $\Delta(1232)$, respectively. Our fits are compatible with these constraints. The LECs $\tilde{d}_{168}$ and $g_E$, appearing in higher-order Lagrangians, are partially constrained as well\footnote {Goldberger-Treiman relation and nucleon-to-$\Delta$ REM relation, respectively.}. However, we find that they are sensitive to higher-order corrections to the description of the process studied here.

\begin{acknowledgments}
 This research was supported by the Spanish Ministerio de Econom\'ia y Competitividad and European FEDER funds under Contracts No. FIS2011-28853-C02-01, FIS2014-51948-C2-2-P and SEV-2014-0398, and the Generalitat Valenciana in the program Prometeo II-2014/068. A.N. Hiller Blin acknowledges support from the Santiago Grisol\'ia program of the  Generalitat Valenciana. We thank D. Hornidge
for providing us with the full set of data from Ref.~\cite{Hornidge:2012ca}.

\end{acknowledgments}


\appendix

\section{The $\Delta(1232)$ loop contribution to the nucleon self energy}\label{AWfrDel}

The expression for $Z_p^\Delta$ coming from the $\Delta$ loop in Fig.~\ref{FSelfEn} is given by
\begin{align}
\nn \frac{h_A^2}{768 \pi ^2 F_{\pi }^2 m^4 M_{\Delta }^2} \Bigg\{&48 m_{\pi }^2 m^5 \left(m+M_{\Delta }\right) \log \left(\frac{m}{M_{\Delta }}\right)\\
\nn&-\left(m-M_{\Delta }\right){}^2
   \left(m+M_{\Delta }\right){}^4 \left(5 m^2-2 m M_{\Delta }+3 M_{\Delta }^2\right) \log \left(\frac{\left(m^2-M_{\Delta
   }^2\right){}^2}{M_{\Delta }^4}\right)\\
\nn &+2 \bigg(5 m^8+8 m^7 M_{\Delta }-6 m^6 \left(2 m_{\pi }^2+M_{\Delta }^2\right)-12
   m^5 M_{\Delta } \left(m_{\pi }^2+M_{\Delta }^2\right)-6 m_{\pi }^4 m^4\\
\nn&-2 m^2 \left(-3 m_{\pi }^4 M_{\Delta }^2+2 m_{\pi
   }^6+M_{\Delta }^6\right)+4 m M_{\Delta } \left(M_{\Delta }^2-m_{\pi }^2\right){}^3+3 \left(m_{\pi }^2-M_{\Delta
   }^2\right){}^4\bigg) \log \left(\frac{m_{\pi }}{M_{\Delta }}\right)\\
\nn&-m_{\pi }^2 m^2 \left(2 m^4+8 m^3 M_{\Delta }+m^2 \left(4 M_{\Delta }^2-5 m_{\pi }^2\right)-8 m
   M_{\Delta } \left(m_{\pi }^2-2 M_{\Delta }^2\right)+6 \left(-3 m_{\pi }^2 M_{\Delta }^2+m_{\pi }^4+3 M_{\Delta }^4\right)\right)\\
\nn&-2
   \sqrt{\left(m-m_{\pi }-M_{\Delta }\right) \left(m+m_{\pi }-M_{\Delta }\right)} \left(\left(m-m_{\pi }+M_{\Delta }\right)
   \left(m+m_{\pi }+M_{\Delta }\right)\right){}^{3/2}\\
\nn &\times\left(5 m^4-2 m^3 M_{\Delta }-2 m^2 \left(m_{\pi }^2+M_{\Delta }^2\right)+2 m
   M_{\Delta } \left(M_{\Delta }^2-m_{\pi }^2\right)-3 \left(m_{\pi }^2-M_{\Delta }^2\right){}^2\right)\\
\nn& \times\Bigg[-\arctanh\Bigg(\frac{m^2+m_{\pi }^2-M_{\Delta }^2}{\sqrt{\left(m-m_{\pi }-M_{\Delta }\right) \left(m+m_{\pi }-M_{\Delta }\right)
   \left(m-m_{\pi }+M_{\Delta }\right) \left(m+m_{\pi }+M_{\Delta }\right)}}\Bigg)\\
&-\arctanh\Bigg(\frac{m^2-m_{\pi }^2+M_{\Delta
   }^2}{\sqrt{\left(m-m_{\pi }-M_{\Delta }\right) \left(m+m_{\pi }-M_{\Delta }\right) \left(m-m_{\pi }+M_{\Delta }\right)
   \left(m+m_{\pi }+M_{\Delta }\right)}}\Bigg)\Bigg]\Bigg\}.
\end{align}

As for the correction piece $m_N^\Delta$ to the nucleon mass, it has the following expression:
\begin{align}
\nn \frac{h_A^2}{768 m^3 \pi ^2 F_{\pi }^2 M_{\Delta }^2}
\Bigg\{
 &+m_{\pi }^2 \left(2 m_{\pi }^4-\left(7 m^2+4
   M_{\Delta } m+6 M_{\Delta }^2\right) m_{\pi }^2+2 \left(m-M_{\Delta }\right) \left(m+M_{\Delta }\right){}^3\right) m^2\\
\nn&+2 \left[m^3+2 M_{\Delta } m^2-2 \left(2 m_{\pi }^2+M_{\Delta }^2\right) m-6 M_{\Delta } \left(m_{\pi }^2+M_{\Delta }^2\right)\right] m^5 \log \left(\frac{m_{\pi }^2}{m^2}\right)\\
\nn &+2 \left(\left(m-M_{\Delta }\right){}^2-m_{\pi }^2\right) \left(m-m_{\pi
   }+M_{\Delta }\right){}^2 \\
\nn &\times\sqrt{\left(m-m_{\pi }-M_{\Delta }\right) \left(m+m_{\pi }-M_{\Delta }\right) \left(m-m_{\pi }+M_{\Delta }\right)} \left(m+m_{\pi }+M_{\Delta }\right){}^{5/2} \log \left(\frac{M_{\Delta }}{m}\right)\\
\nn &+2\frac{
   \left(\left(m-M_{\Delta }\right){}^2-m_{\pi }^2\right){}^2 \left(m-m_{\pi }+M_{\Delta }\right){}^3 \left(m+m_{\pi }+M_{\Delta }\right){}^{5/2}}{\sqrt{\left(m-m_{\pi }-M_{\Delta }\right) \left(m+m_{\pi }-M_{\Delta }\right) \left(m-m_{\pi
   }+M_{\Delta }\right)}} \log \left(\frac{m_{\pi }}{m}\right)\\
\nn &-\frac{ \left(\left(m-M_{\Delta
   }\right){}^2-m_{\pi }^2\right){}^2 \left(m-m_{\pi }+M_{\Delta }\right){}^3 \left(m+m_{\pi }+M_{\Delta }\right){}^{5/2}}{\sqrt{\left(m-m_{\pi }-M_{\Delta }\right) \left(m+m_{\pi }-M_{\Delta }\right) \left(m-m_{\pi }+M_{\Delta }\right)}}\\
\nn&\times\log \left(\frac{\left(-m^2+m_{\pi }^2+M_{\Delta }^2+\sqrt{m^4-2 \left(m_{\pi }^2+M_{\Delta }^2\right) m^2+\left(M_{\Delta }^2-m_{\pi }^2\right){}^2}\right){}^2}{4 m^4}\right)\\
\nn&- \left(m-M_{\Delta }\right) \left(m+M_{\Delta }\right){}^3 \left[2 \left(-2 m^2+M_{\Delta } m-2 M_{\Delta }^2\right) m_{\pi }^2+\left(m^2-M_{\Delta }^2\right){}^2\right]\log
   \left(\frac{\left(m^2-M_{\Delta }^2\right){}^2}{m^4}\right)\\
\nn &-2  \bigg[m^8+2 M_{\Delta } m^7-2 \left(2 m_{\pi }^2+M_{\Delta }^2\right) m^6-6 M_{\Delta } \left(m_{\pi }^2+M_{\Delta }^2\right) m^5\\
\nn&+6 m_{\pi }^4 m^4+6 M_{\Delta } \left(m_{\pi }^4-M_{\Delta }^4\right) m^3-2
   \left(2 m_{\pi }^6-3 M_{\Delta }^2 m_{\pi }^4+M_{\Delta }^6\right) m^2\\
\nn&+2 M_{\Delta } \left(M_{\Delta }^2-m_{\pi }^2\right){}^3 m+\left(M_{\Delta }^2-m_{\pi }^2\right){}^4\bigg]\log
   \left(\frac{m_{\pi }}{m}\right)\\
\nn &+2\bigg[m_{\pi }^8-2
   \left(2 m^2+M_{\Delta } m+2 M_{\Delta }^2\right) m_{\pi }^6+6 \left(m^4+M_{\Delta } m^3+M_{\Delta }^2 m^2+M_{\Delta }^3 m+M_{\Delta }^4\right) m_{\pi }^4\\
&+2 \left(-2 m^6-3 M_{\Delta } m^5+3 M_{\Delta }^5 m+2 M_{\Delta }^6\right) m_{\pi
   }^2+\left(m-M_{\Delta }\right){}^3 \left(m+M_{\Delta }\right){}^5\bigg] \log \left(\frac{M_{\Delta }}{m}\right) \Bigg\}.
\end{align}

\section{Diagrams' amplitudes}\label{ADiagrams}

\begin{figure}[htb]
\begin{center}
\subfigure[]{
\label{fO3LMa}
\includegraphics[width=0.3\textwidth]{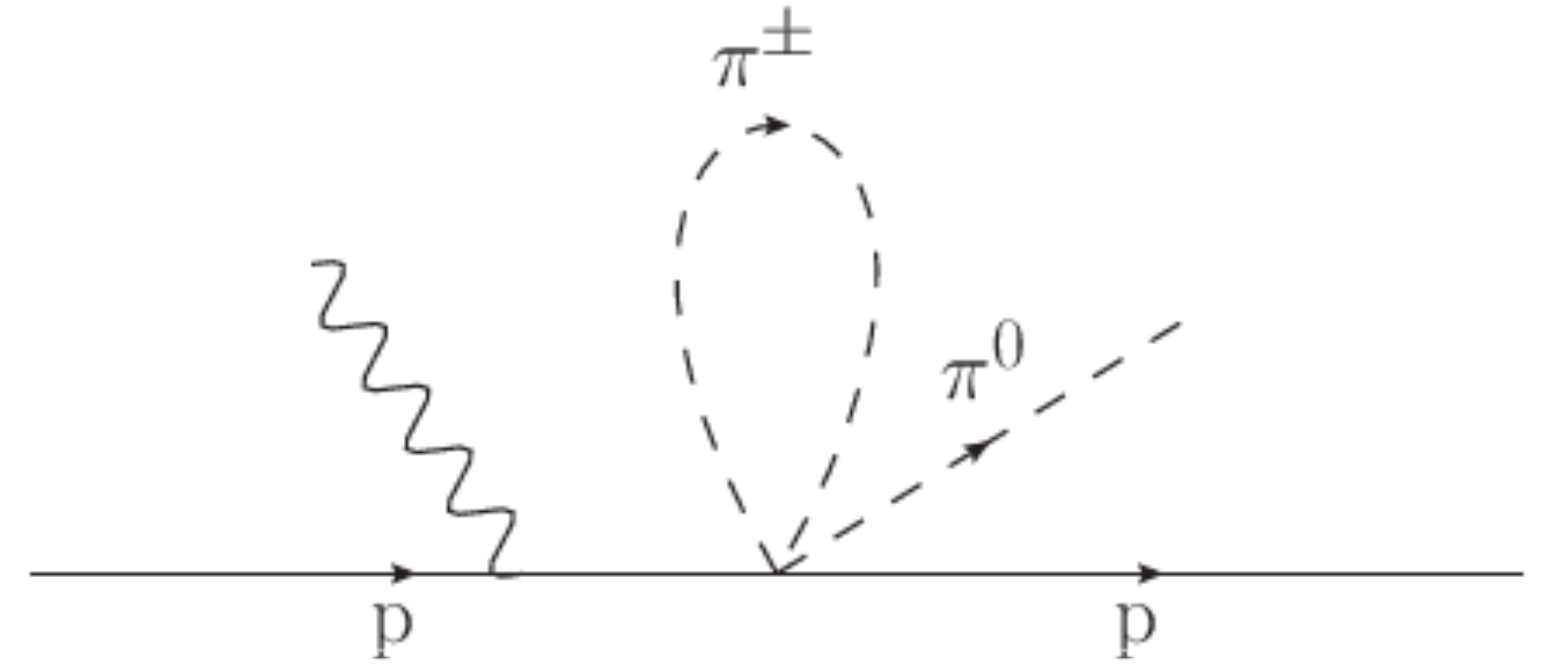}}
\subfigure[]{
\label{fO3LMc}
\includegraphics[width=0.3\textwidth]{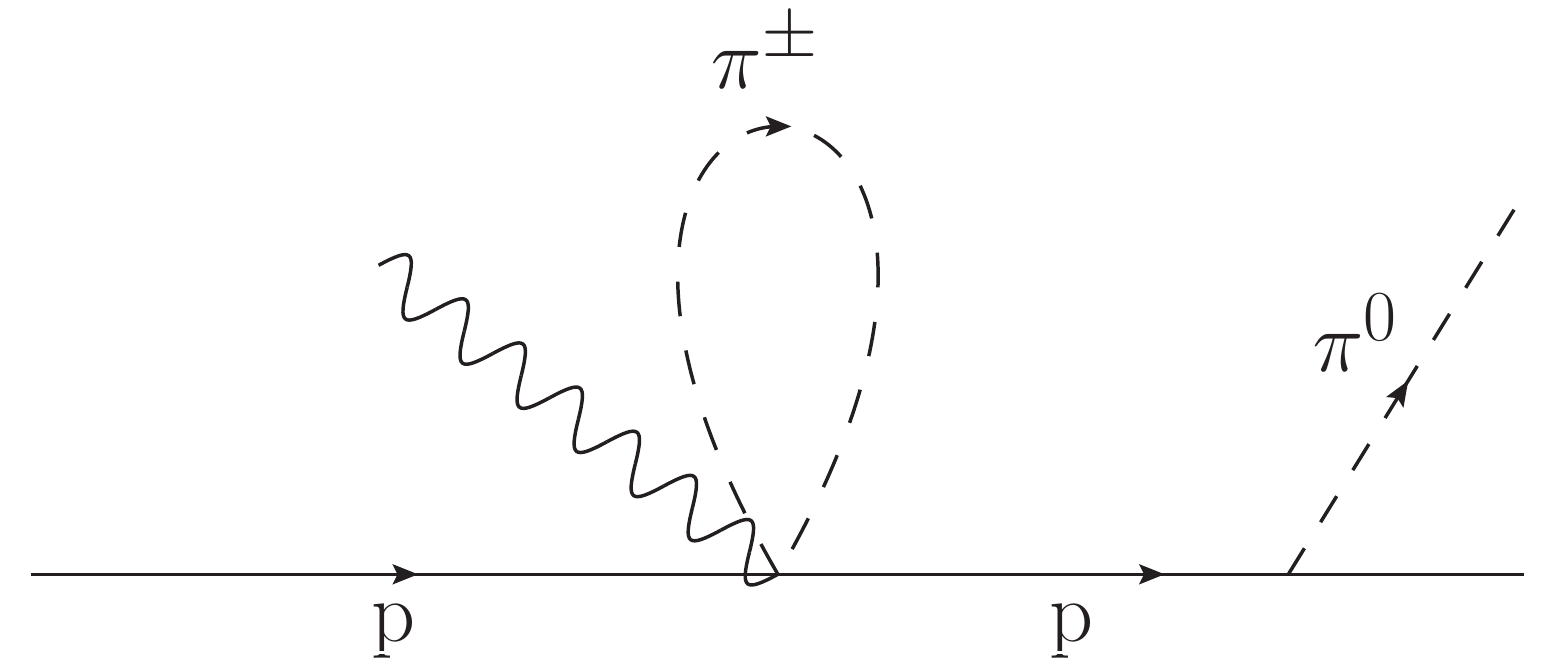}}
\subfigure[]{
\label{fO3LMe}
\includegraphics[width=0.3\textwidth]{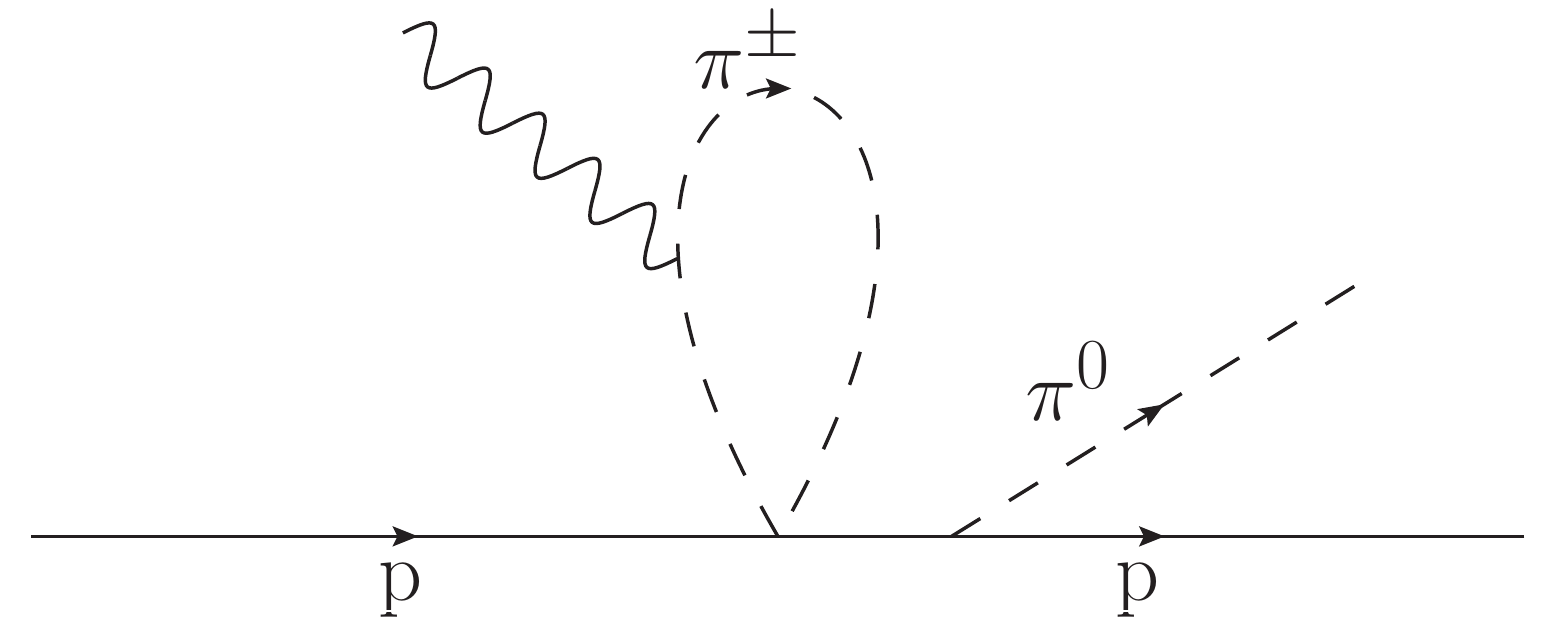}}\\
\subfigure[]{
\label{fO3LB1a}
\includegraphics[width=0.3\textwidth]{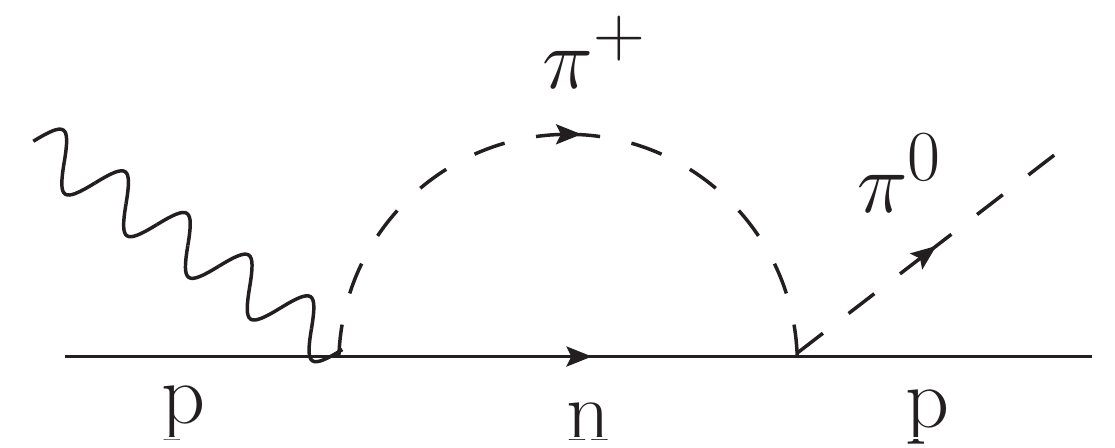}}
\subfigure[]{
\label{fO3LB1b}
\includegraphics[width=0.3\textwidth]{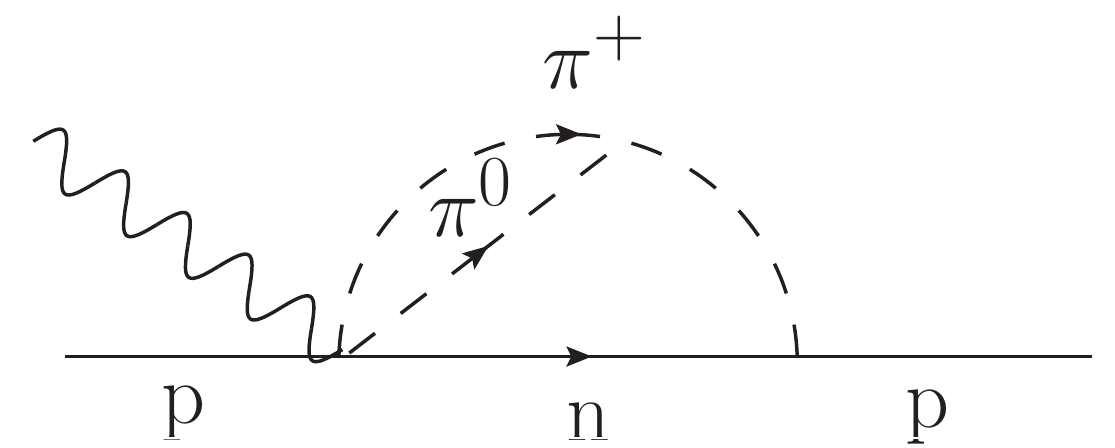}}
\subfigure[]{
\label{fO3LB1e}
\includegraphics[width=0.3\textwidth]{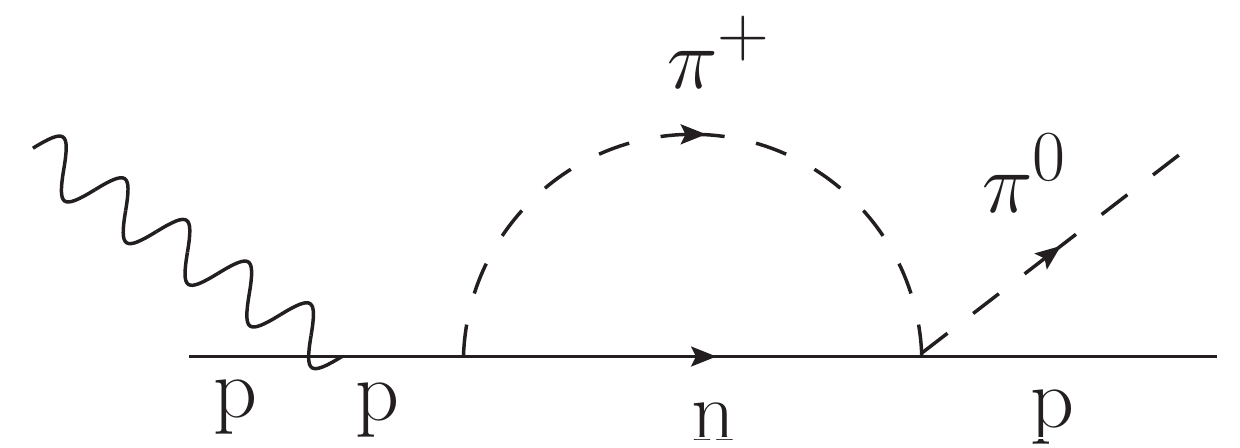}}\\
\subfigure[]{
\label{fO3LB1i}
\includegraphics[width=0.3\textwidth]{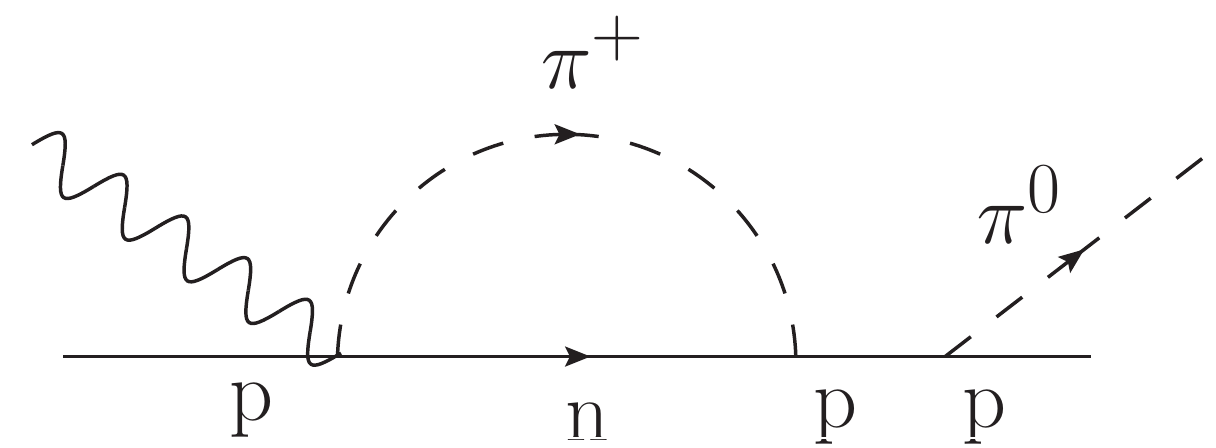}}
\subfigure[]{
\label{fO3LB1m}
\includegraphics[width=0.3\textwidth]{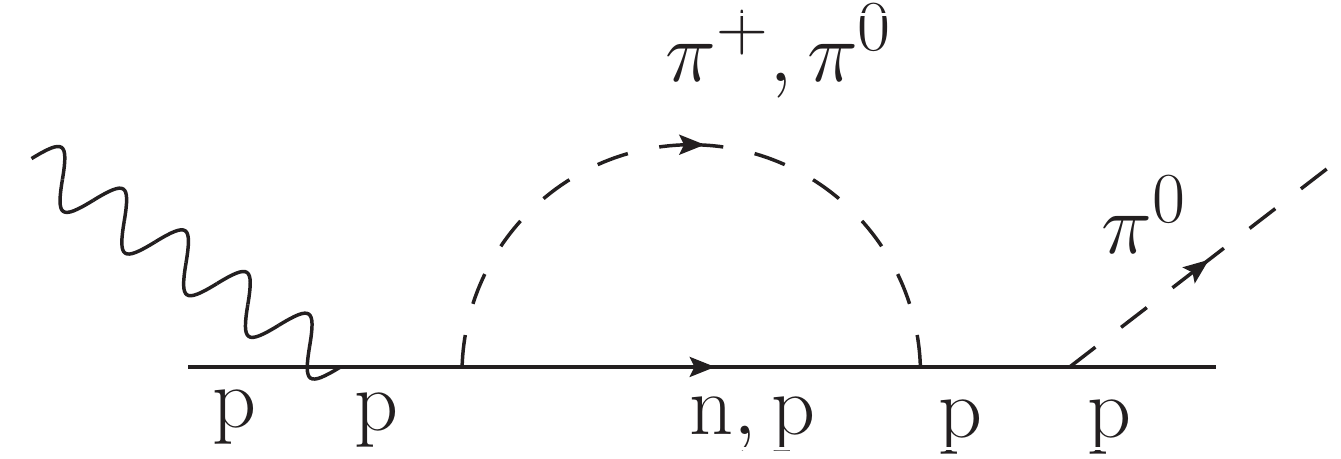}}
\subfigure[]{
\label{fO3LBM3a}
\includegraphics[width=0.3\textwidth]{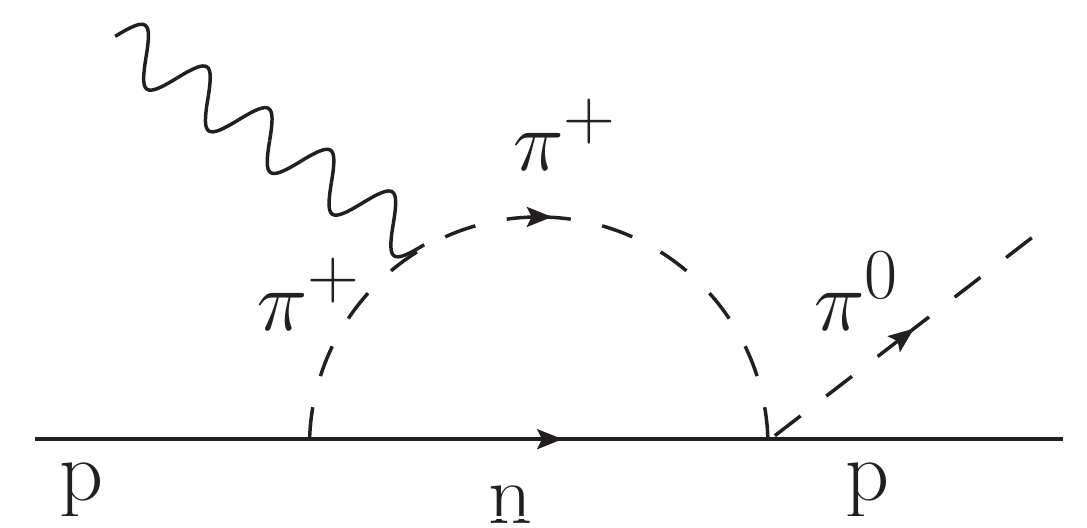}}\\
\subfigure[]{
\label{fO3LBM3c}
\includegraphics[width=0.3\textwidth]{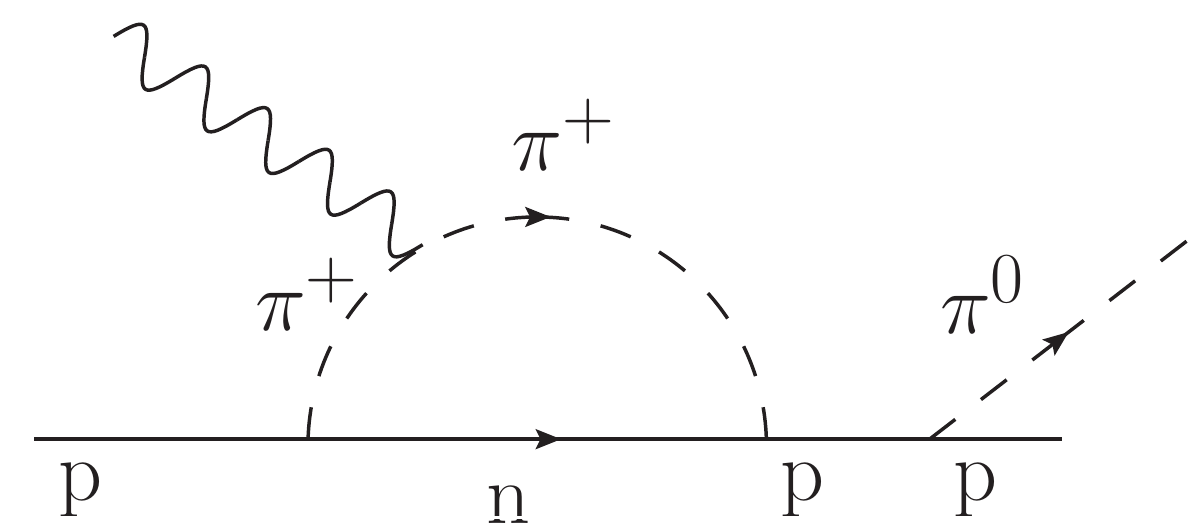}}
\subfigure[]{
\label{fO3LBM3e}
\includegraphics[width=0.3\textwidth]{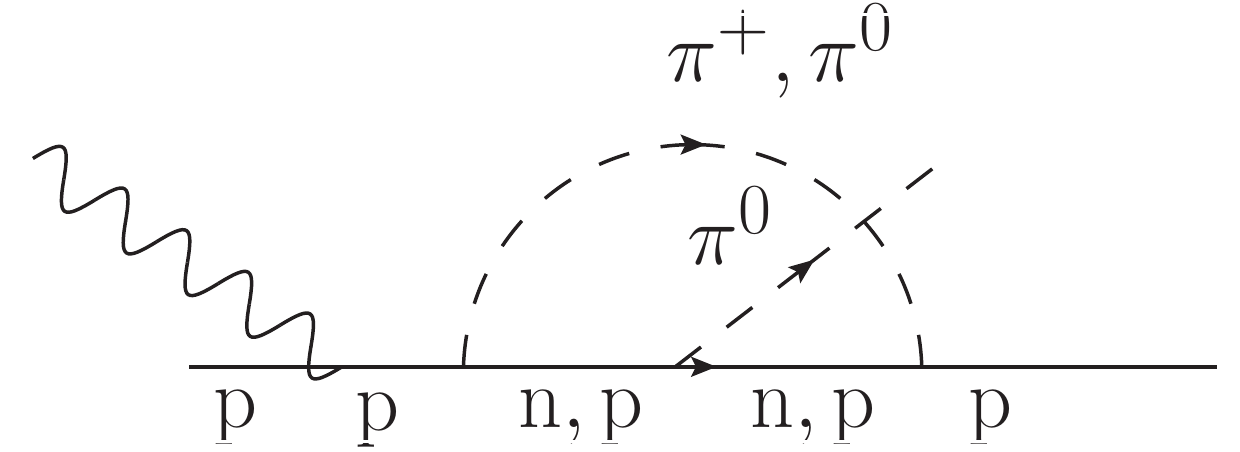}}
\subfigure[]{
\label{fO3LBM3g}
\includegraphics[width=0.3\textwidth]{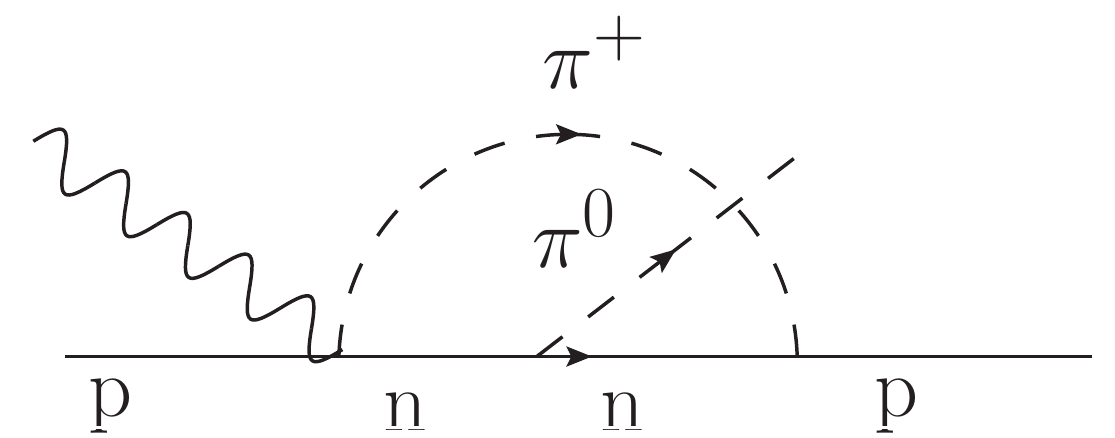}}\\
\subfigure[]{
\label{fO3LBM3i}
\includegraphics[width=0.3\textwidth]{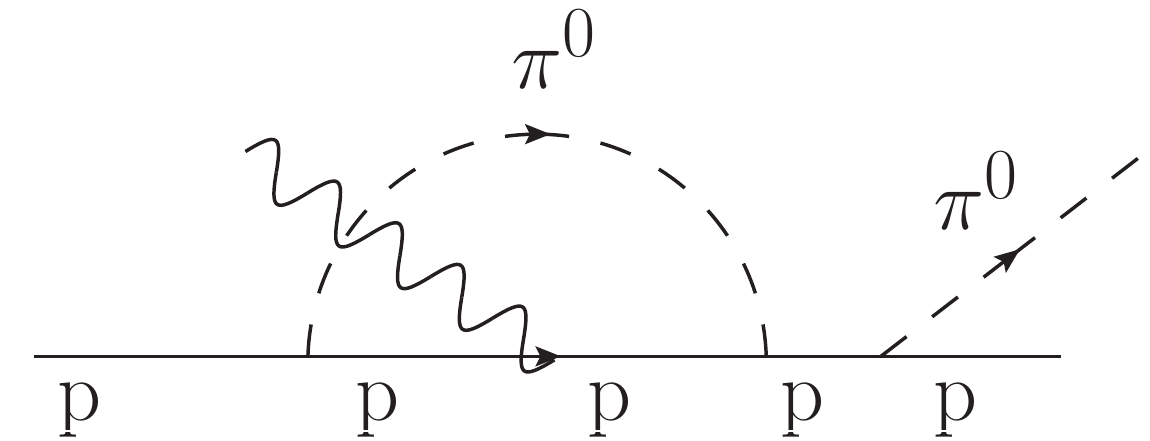}}
\subfigure[]{
\label{fO3LBM4a}
\includegraphics[width=0.3\textwidth]{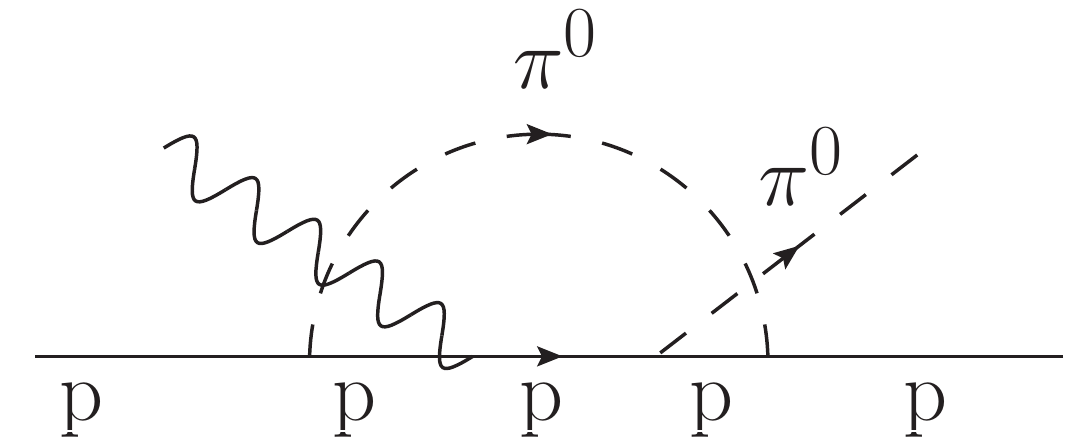}}
\subfigure[]{
\label{fO3LBM4b}
\includegraphics[width=0.3\textwidth]{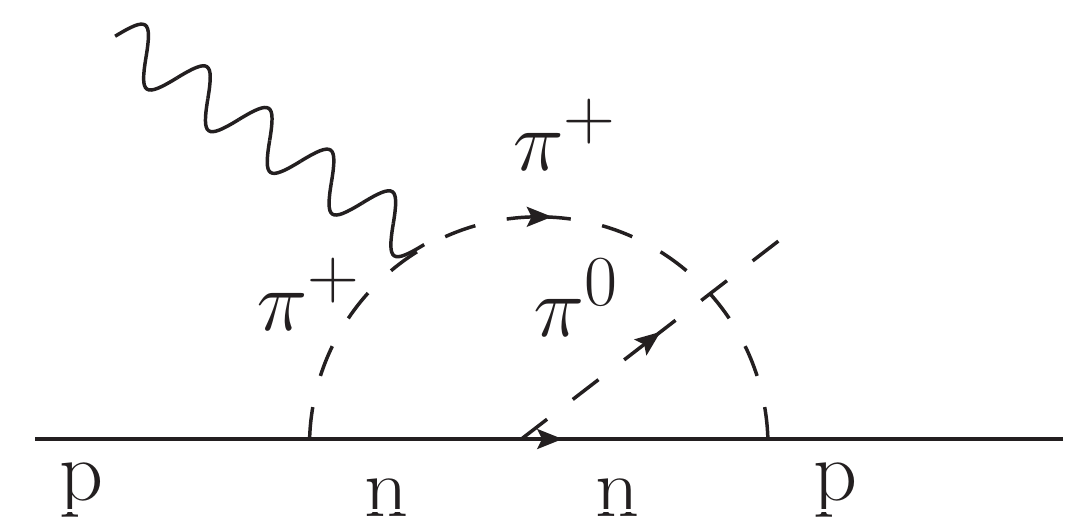}}
\end{center}
\caption{Loop diagrams for $\pi^0$ photoproduction off protons, including only nucleonic intermediate states. The crossed terms are not shown, but also calculated.}\label{FLoop12}
\end{figure}

\begin{figure}[htb]
\begin{center}
\subfigure[]{
\label{fD1a}
\includegraphics[width=0.25\textwidth]{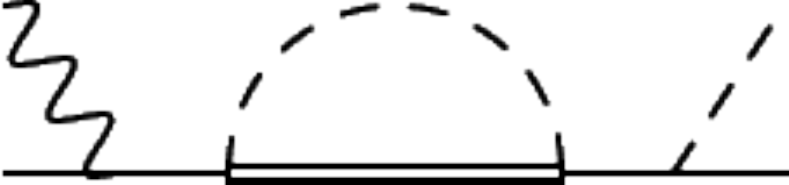}}
\subfigure[]{
\label{fD1b}
\includegraphics[width=0.25\textwidth]{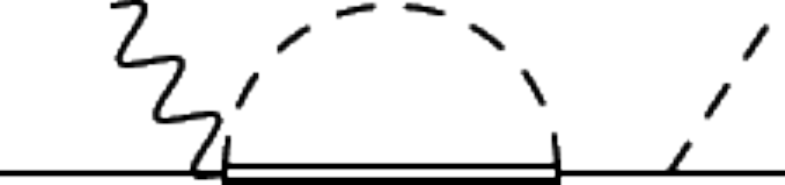}}
\subfigure[]{
\label{fD1c}
\includegraphics[width=0.25\textwidth]{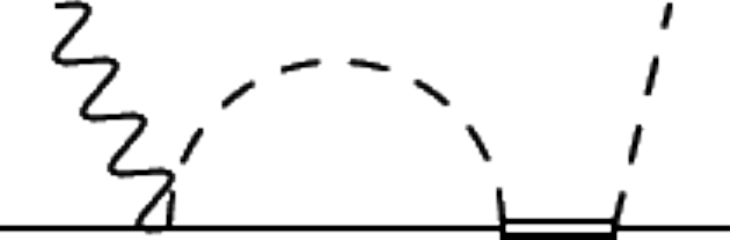}}\\
\subfigure[]{
\label{fD1d}
\includegraphics[width=0.25\textwidth]{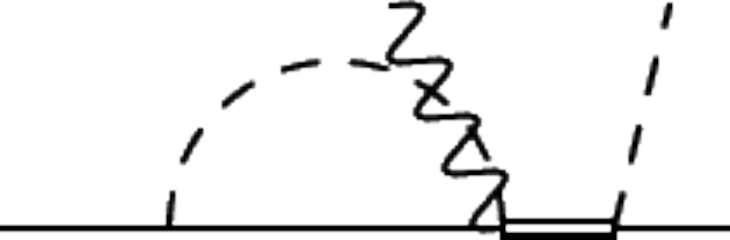}}
\subfigure[]{
\label{fD1e}
\includegraphics[width=0.25\textwidth]{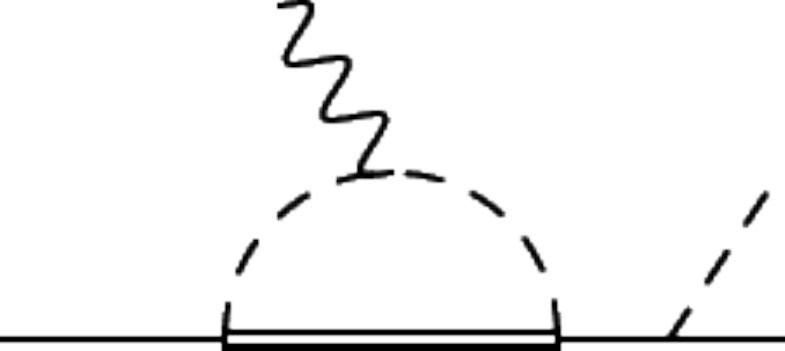}}
\subfigure[]{
\label{fD1f}
\includegraphics[width=0.25\textwidth]{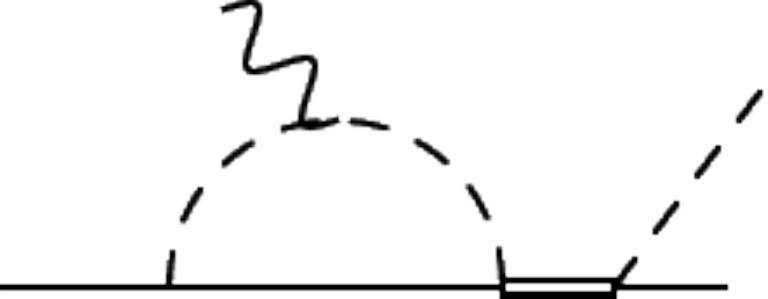}}\\
\subfigure[]{
\label{fD1g}
\includegraphics[width=0.25\textwidth]{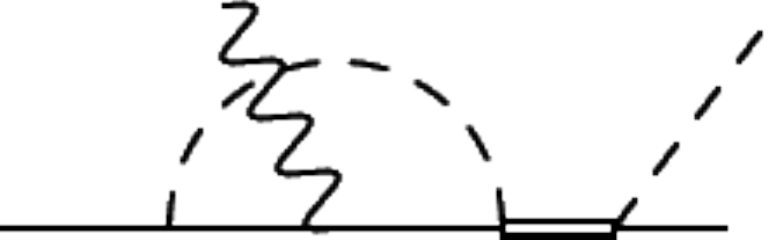}}
\subfigure[]{
\label{fD1h}
\includegraphics[width=0.25\textwidth]{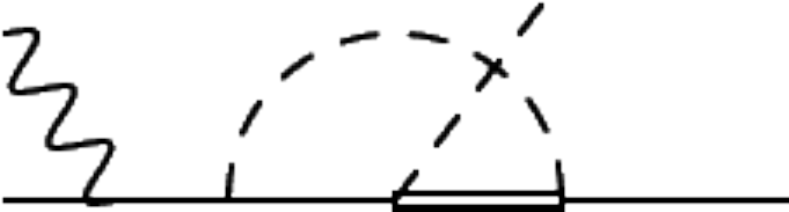}}
\subfigure[]{
\label{fD1i}
\includegraphics[width=0.25\textwidth]{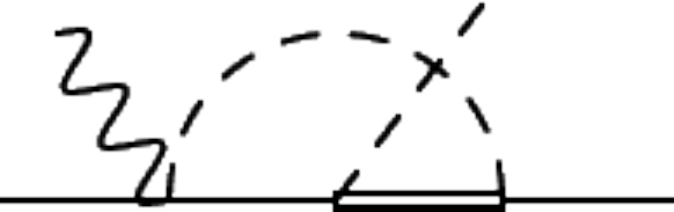}}\\
\subfigure[]{
\label{fD1j}
\includegraphics[width=0.25\textwidth]{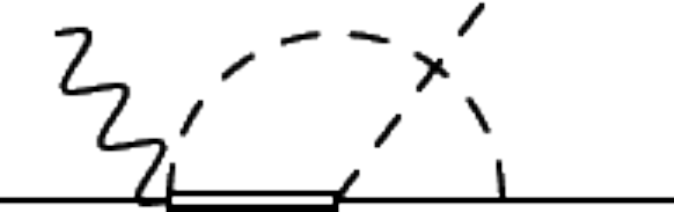}}
\subfigure[]{
\label{fD1k}
\includegraphics[width=0.25\textwidth]{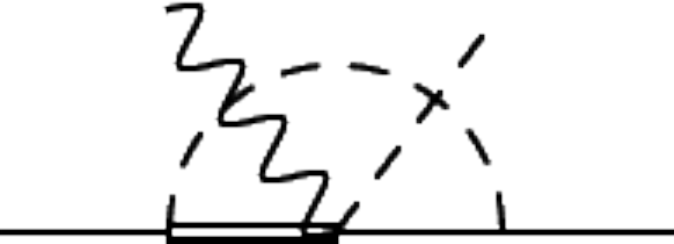}}
\subfigure[]{
\label{fD1l}
\includegraphics[width=0.25\textwidth]{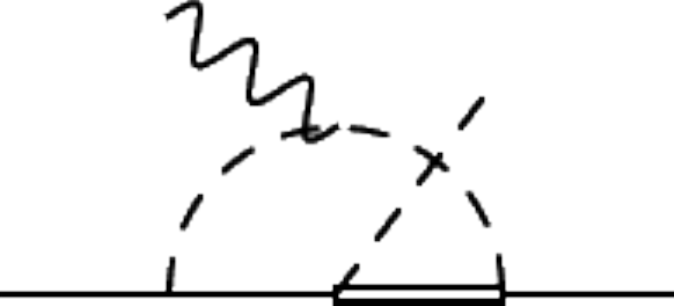}}\\
\subfigure[]{
\label{fD1m}
\includegraphics[width=0.25\textwidth]{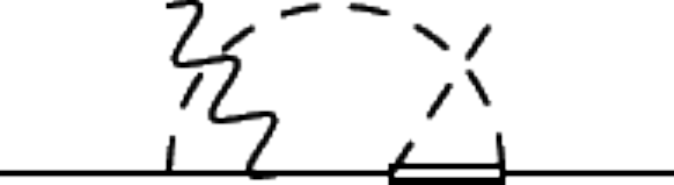}}
\subfigure[]{
\label{fD1n}
\includegraphics[width=0.25\textwidth]{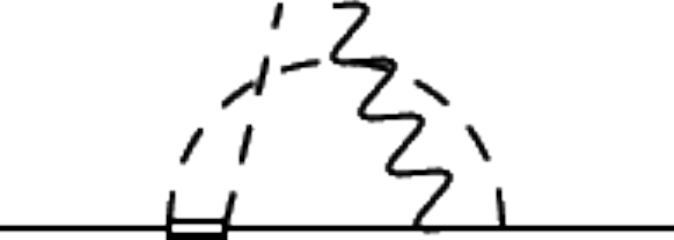}}
\end{center}
\caption{Loop diagrams for $\pi^0$ photoproduction off protons for $\Delta$ intermediate states. The crossed terms
are not shown, but also calculated.}\label{FLoop32}
\end{figure}

The tree-level diagrams of Fig.~\ref{FTree12} have the following amplitude expressions:
\begin{align}
\nn \mathcal{M}_{\ref{fO1a},p^1}=&-\mathrm{i}\frac{eg_0}{2F}\bra{p',q}
\Bigg[\left(\frac{m+m_2}{m^2+2p\cdot k-m_2^2}+\frac{m+m_2}{m^2-2p'\cdot k-m_2^2}\right)\sle\slk\gamma_5\\
&+\left(\frac{2p\cdot k}{m^2+2p\cdot k-m_2^2}+\frac{2p'\cdot k}{m^2-2p'\cdot k-m_2^2}\right)\sle\gamma_5
+\frac{2(m+m_2)}{m^2-2p'\cdot k-m_2^2} \epsilon\cdot q\gamma_5\Bigg]
\ket{p},\\[1em]
\mathcal{M}_{\ref{fO1a},p^2}=&-\mathrm{i}\frac{eg_A}{8mF_\pi}\left(c_6+c_7\right)\bra{p',q}
\left[\left(4+4m^2\left(\frac{1}{p\cdot k}-\frac{1}{p'\cdot k}\right)\right)\sle\slk\gamma_5
-\frac{4m}{p'\cdot k}\epsilon\cdot q\slk\gamma_5\right]
\ket{p},\\[1em]
 \mathcal{M}_{\ref{fO1a},p^3} =&\frac{\mathrm{i}e(2d_{16}-d_{18})m~m_\pi^2}{F_\pi}\bra{p',q}
 \left[\left(\frac{1}{p'\cdot k}-\frac{1}{p\cdot k}\right)\sle\slk\gamma_5 + \frac{2}{p'\cdot k} \epsilon\cdot q\gamma_5 \right]
\ket{p},\\[1em]
 \mathcal{M}_{\ref{fO1c}} =&\frac{4\mathrm{i}e(d_8+d_9)}{F_\pi}\bra{p',q}
 \left[\frac{1}{m}\big(m^2+p\cdot p'\big)\sle\slk\gamma_5+\frac{1}{m}p\cdot k \epsilon\cdot q\gamma_5+(p+p')\cdot k\sle\gamma_5 + \epsilon\cdot q\slk\gamma_5\right]
\ket{p}.
\end{align}
Note that the nucleon mass $m$ is set to the physical nucleon mass $m_N$ everywhere except in the propagator of the $\mathcal{O}(p^1)$ amplitude, where we perform the correction shown in Eq.~\ref{eq:17}.

The amplitudes of the diagrams of Figs.~\ref{fO3LMa} to~\ref{fO3LMe} combined have the following simple expression. The sum over isospin channels has already been performed:
\begin{align}
\mathcal{M}_{\ref{fO3LMa},\ref{fO3LMc},\ref{fO3LMe}} =&\frac{\mathrm{i}emm_\pi^2g_A}{96\pi^2F_\pi^3}\left(\lambda-\log\left[\frac{m_\pi^2}{m^2}\right]\right)\bra{p',q}
\left[\left(\frac{1}{p'\cdot k}-\frac{1}{p\cdot k}\right)\sle\slk\gamma_5
+\frac{2}{p'\cdot k}\epsilon\cdot q\gamma_5\right]
\ket{p},
\end{align}
where $\lambda=\frac{2}{\epsilon}+\log(4\pi)-\gamma_E+1+\mathcal{O}(\epsilon)$ is the piece that is EOMS-renormalized according to the $\widetilde{MS}$ scheme. Note that we are using the nucleon mass $m$ as the chiral-symmetry breaking scale. As for the other Fig.~\ref{FLoop12} diagrams' expressions, they are listed here before being evaluated, as they have rather large expressions:
\begin{align}
\nn \mathcal{M}_{\ref{fO3LB1a}} =\frac{eg_A}{4F_\pi^3}
\int{\frac{\mathrm{d}^dz}{(2\pi)^d}}
\bra{p',q}&\frac{(\slz+\slq)(\slp+\slk-\slz+m)\sle\gamma_5}{[z^2-m_\pi^2][(z-p-k)^2-m^2]}\\
 - &\frac{\sle\gamma_5(\slp-\slq-\slz+m)(\slz-\slq)}{[z^2-m_\pi^2][(z-p+q)^2-m^2]}\ket{p}.
\end{align}
The contributions of the direct and crossed diagrams in Fig.~\ref{fO3LB1b} exactly cancel each other.
\begin{align}
\nn \mathcal{M}_{\ref{fO3LB1e}} =&\frac{eg_A}{4F_\pi^3}
\int{\frac{\mathrm{d}^dz}{(2\pi)^d}}
\bra{p',q}\\
\nn &-\frac{(\slz+\slq)(\slp+\slk-\slz+m)\slz\gamma_5(\slp + \slk +m)\sle}
{[z^2-m_\pi^2][(z-p-k)^2-m^2]2p\cdot k}\\
\nn &+\frac{\slz\gamma_5(\slp'-\slz+m)(\slz-\slq)(\slp + \slk +m)\sle}
{[z^2-m_\pi^2][(z-p')^2-m^2]2p\cdot k}\\
\nn &-\frac{\sle(\slp' - \slk +m)\slz\gamma_5(\slp'-\slk-\slz+m)(\slz-\slq)}
{[z^2-m_\pi^2][(z-p'+k)^2-m^2]2p'\cdot k}\\
 &+\frac{\sle(\slp' - \slk +m)(\slz+\slq)(\slp-\slz+m)\slz\gamma_5}
{[z^2-m_\pi^2][(z-p)^2-m^2]2p'\cdot k}
\ket{p},\\[1em]
\nn \mathcal{M}_{\ref{fO3LB1i}} =&\frac{eg_A^3}{4F_\pi^3}
\int{\frac{\mathrm{d}^dz}{(2\pi)^d}}
\bra{p',q}\\
\nn &-\frac{\slq\gamma_5(\slp + \slk +m)\slz\gamma_5(\slp+\slk-\slz+m)\sle\gamma_5}
{[z^2-m_\pi^2][(z-p-k)^2-m^2]2p\cdot k}\\
\nn &+\frac{\slz\gamma_5(\slp'-\slz+m)\sle\gamma_5(\slp' - \slk +m)\slq\gamma_5}
{[z^2-m_\pi^2][(z-p')^2-m^2]2p'\cdot k}\\
\nn &+\frac{\sle\gamma_5(\slp'-\slk-\slz+m)\slz\gamma_5(\slp' - \slk +m)\slq\gamma_5}
{[z^2-m_\pi^2][(z-p'+k)^2-m^2]2p'\cdot k}\\
 &-\frac{\slq\gamma_5(\slp + \slk +m)\sle\gamma_5(\slp-\slz+m)\slz\gamma_5}
{[z^2-m_\pi^2][(z-p)^2-m^2]2p\cdot k}
\ket{p},\\[1em]
\nn \mathcal{M}_{\ref{fO3LB1m}} =&\frac{3eg_A^3}{8F_\pi^3}
\int{\frac{\mathrm{d}^dz}{(2\pi)^d}}
\bra{p',q}\\
\nn &\frac{\slq\gamma_5(\slp + \slk +m)\slz\gamma_5(\slp+\slk-\slz+m)\slz\gamma_5(\slp + \slk +m)\sle}
{[z^2-m_\pi^2][(z-p-k)^2-m^2]4(p\cdot k)^2}\\
\label{eqbarloopmn} &+\frac{\sle(\slp' - \slk +m)\slz\gamma_5(\slp'-\slk-\slz+m)\slz\gamma_5(\slp' - \slk +m)\slq\gamma_5}
{[z^2-m_\pi^2][(z-p'+k)^2-m^2]4(p'\cdot k)^2}
\ket{p},\\[1em]
\nn \mathcal{M}_{\ref{fO3LBM3a}} =&\frac{eg_A}{2F_\pi^3}
\int{\frac{\mathrm{d}^dz}{(2\pi)^d}}
\bra{p',q}\\
\nn &-\frac{(\slz+\slq+\slk)\epsilon\cdot z(\slp-\slz +m)\slz\gamma_5}
{[z^2-m_\pi^2][(z-p)^2-m^2][(z+k)^2-m_\pi^2]}\\
 &+\frac{(\slz+\slk)\gamma_5\epsilon\cdot z(\slp'-\slk-\slz +m)(\slz-\slq)}
{[z^2-m_\pi^2][(z-p'+k)^2-m^2][(z+k)^2-m_\pi^2]}
\ket{p}\\[1em]
\nn \mathcal{M}_{\ref{fO3LBM3c}} =&\frac{eg_A^3}{2F_\pi^3}
\int{\frac{\mathrm{d}^dz}{(2\pi)^d}}
\bra{p',q}\\
\nn &\frac{\slq\gamma_5(\slp+\slk+m)(\slz+\slk)\gamma_5\epsilon\cdot z(\slp-\slz +m)\slz\gamma_5}
{[z^2-m_\pi^2][(z-p)^2-m^2][(z+k)^2-m_\pi^2]2p\cdot k}\\
 &-\frac{(\slz+\slk)\gamma_5\epsilon\cdot z(\slp'-\slk-\slz +m)\slz\gamma_5(\slp'-\slk+m)\slq\gamma_5}
{[z^2-m_\pi^2][(z-p'+k)^2-m^2][(z+k)^2-m_\pi^2]2p'\cdot k}
\ket{p},\\[1em]
\nn \mathcal{M}_{\ref{fO3LBM3e}} =&\frac{eg_A^3}{8F_\pi^3}
\int{\frac{\mathrm{d}^dz}{(2\pi)^d}}
\bra{p',q}\\
\nn &-\frac{\slz\gamma_5(\slp'-\slz +m)\slq\gamma_5(\slp+\slk-\slz+m)\slz\gamma_5(\slp+\slk +m)\sle}
{[z^2-m_\pi^2][(z-p')^2-m^2][(z-p-k)^2-m^2]2p\cdot k}\\
 &+\frac{\sle(\slp'-\slk+m)\slz\gamma_5(\slp'-\slk-\slz+m)\slq\gamma_5(\slp-\slz +m)\slz\gamma_5}
{[z^2-m_\pi^2][(z-p'+k)^2-m^2][(z-p)^2-m^2]2p'\cdot k}
\ket{p},\\[1em]
\nn \mathcal{M}_{\ref{fO3LBM3g}} =&\frac{eg_A^3}{4F_\pi^3}
\int{\frac{\mathrm{d}^dz}{(2\pi)^d}}
\bra{p',q}\\
\nn &\frac{\slz\gamma_5(\slp'-\slz +m)\slq\gamma_5(\slp+\slk-\slz+m)\sle\gamma_5}
{[z^2-m_\pi^2][(z-p')^2-m^2][(z-p-k)^2-m^2]}\\
 +&\frac{\sle\gamma_5(\slp'-\slk-\slz+m)\slq\gamma_5(\slp-\slz +m)\slz\gamma_5}
{[z^2-m_\pi^2][(z-p'+k)^2-m^2][(z-p)^2-m^2]}
\ket{p},\\[1em]
\nn \mathcal{M}_{\ref{fO3LBM3i}} =&\frac{eg_A^3}{8F_\pi^3}
\int{\frac{\mathrm{d}^dz}{(2\pi)^d}}
\bra{p',q}\\
\nn &\frac{\slq\gamma_5(\slp+\slk+m)\slz\gamma_5(\slp+\slk-\slz+m)\sle(\slp-\slz+m)\slz\gamma_5}
{[z^2-m_\pi^2][(z-p)^2-m^2][(z-p-k)^2-m^2]2p\cdot k}\\
 &-\frac{\slz\gamma_5(\slp'-\slz+m)\sle(\slp'-\slk-\slz +m)\slz\gamma_5(\slp'-\slk+m)\slq\gamma_5}
{[z^2-m_\pi^2][(z-p'+k)^2-m^2][(z-p')^2-m^2]2p'\cdot k}
\ket{p},\\[1em]
\nn \mathcal{M}_{\ref{fO3LBM4a}} =&\frac{eg_A^3}{8F_\pi^3}
\int{\frac{\mathrm{d}^dz}{(2\pi)^d}}
\bra{p',q}\\
\nn &\frac{\slz\gamma_5(\slp'-\slz +m)\slq\gamma_5(\slp+\slk-\slz+m)\sle(\slp-\slz+m)\slz\gamma_5}
{[z^2-m_\pi^2][(z-p')^2-m^2][(z-p-k)^2-m^2][(z-p)^2-m^2]}\\
 &+\frac{\slz\gamma_5(\slp'-\slz+m)\sle(\slp'-\slk-\slz+m)\slq\gamma_5(\slp-\slz +m)\slz\gamma_5}
{[z^2-m_\pi^2][(z-p')^2-m^2][(z-p'+k)^2-m^2][(z-p)^2-m^2]}\ket{p},\\[1em]
\nn \mathcal{M}_{\ref{fO3LBM4b}} =&-\frac{eg_A^3}{2F_\pi^3}
\int{\frac{\mathrm{d}^dz}{(2\pi)^d}}
\bra{p',q}\\
 &\frac{(\slz+\slk)\gamma_5\epsilon\cdot z(\slp'-\slk-\slz+m)\slq\gamma_5(\slp-\slz+m)\slz\gamma_5}
{[z^2-m_\pi^2][(z+k)^2-m_\pi^2][(z-p)^2-m^2][(z-p'+k)^2-m^2]}\ket{p}.
\end{align}

As for the diagrams including $\Delta$ propagators, we introduce the definition
\begin{align}
\nn S_\Delta^{\alpha\beta}(p)=
\frac{\slp+M_\Delta}{p^2-M_\Delta^2+\mathrm{i}\varepsilon}
\left[
-g^{\alpha\beta}
+ \frac{1}{D-1}\gamma^\alpha\gamma^\beta
+ \frac{1}{(D-1)M_\Delta}(\gamma^\alpha p^\beta - \gamma^\beta p^\alpha)
+ \frac{D-2}{(D-1)M_\Delta^2}p^\alpha p^\beta
\right]
\end{align}
for simplicity, where $D$ is the Minkowski-space dimension. The tree-level amplitudes of Fig.~\ref{FTree32} then read:
\begin{align}
\nn\mathcal{M}_{\ref{fDd},p^2}=&\bra{p',q}-\frac{eh_Ag_M}{2mM_\Delta(m+M_\Delta)F_\pi}
\Big[(p_\mu+k_\mu)q_\lambda\gamma^{\mu\nu\lambda}S_\Delta^{\nu\beta}(p+k)
(p_\alpha + k_\alpha)k_\delta\epsilon_\rho\epsilon^{\alpha\beta\delta\rho}\\
&-(p'_\alpha - k_\alpha)k_\delta\epsilon_\rho\epsilon^{\alpha\beta\delta\rho}
S_\Delta^{\beta\nu}(p'-k)(p'_\mu-k_\mu)q_\lambda\gamma^{\mu\nu\lambda}\Big]
\ket{p},\\[1em]
\nn\mathcal{M}_{\ref{fDd},p^3}=&\bra{p',q}-\mathrm{i}\frac{eh_Ag_E}{2mM_\Delta(m+M_\Delta)F_\pi}
\Big[(p_\mu+k_\mu)q_\lambda\gamma^{\mu\nu\lambda}S_\Delta^{\nu\beta}(p+k)
(p_\alpha + k_\alpha)(k^\alpha\epsilon^\beta-k^\beta\epsilon^\alpha)\gamma_5\\
&-(p'_\alpha - k_\alpha)(k^\alpha\epsilon^\beta-k^\beta\epsilon^\alpha)\gamma_5
S_\Delta^{\beta\nu}(p'-k)(p'_\mu-k_\mu)q_\lambda\gamma^{\mu\nu\lambda}
\Big]\ket{p}.
\end{align}
The loop diagrams of Fig.~\ref{FLoop32} also have very large expressions after evaluation and therefore we opt to show the expressions before momentum integration and action of the Dirac equation. The sum over the isospin channels was already performed:
\begin{align}
\nn \mathcal{M}_{\ref{fD1a}} =-&\frac{eg_Ah_A^2}{16F_\pi^3M_{\Delta}^2}
\int{\frac{\mathrm{d}^dz}{(2\pi)^d}}
\bra{p',q}\Bigg\{\\
\nn &\frac{\slq\gamma_5(\slp+\slk+m)(p+k-z)_\alpha z_\delta\gamma^{\alpha\beta\delta} S_\Delta^{\beta\nu}(p+k-z)
(p+k-z)_\mu z_\lambda\gamma^{\mu\nu\lambda}(\slp+\slk+m)\sle}{[z^2-m_\pi^2](p\cdot k)^2}\\
&+\frac{\sle(\slp'-\slk+m)(p'-k-z)_\alpha z_\delta\gamma^{\alpha\beta\delta} S_\Delta^{\beta\nu}(p'-k-z)(p'-k-z)_\mu z_\lambda\gamma^{\mu\nu\lambda}(\slp'-\slk+m)\slq\gamma_5}{[z^2-m_\pi^2](p'\cdot k)^2}\Bigg\}
\ket{p},\\[1em]
\nn \mathcal{M}_{\ref{fD1b}} =&-\frac{eg_Ah_A^2}{24F_\pi^3M_{\Delta}^2}
\int{\frac{\mathrm{d}^dz}{(2\pi)^d}}
\bra{p',q}\Bigg\{\\
\nn&\frac{\slq\gamma_5(\slp+\slk+m)(p+k-z)_\alpha z_\delta\gamma^{\alpha\beta\delta} S_\Delta^{\beta\nu}(p+k-z)(p+k-z)_\mu\epsilon_\lambda
\gamma^{\mu\nu\lambda}}{[z^2-m_\pi^2]p\cdot k}\\
\nn&+\frac{(p'-k-z)_\alpha\epsilon_\delta
\gamma^{\alpha\beta\delta}S_\Delta^{\beta\nu}(p'-k-z) 
(p'-k-z)_\mu z_\lambda\gamma^{\mu\nu\lambda}(\slp'-\slk+m)\slq\gamma_5}{[z^2-m_\pi^2]p'\cdot k}\\
\nn&-\frac{\slq\gamma_5(\slp+\slk+m)(p-z)_\alpha\epsilon_\delta
\gamma^{\alpha\beta\delta} 
 S_\Delta^{\beta\nu}(p-z)(p-z)_\mu z_\lambda\gamma^{\mu\nu\lambda}}{[z^2-m_\pi^2]p\cdot k}\\
&+\frac{(p'-z)_\alpha z_\delta\gamma^{\alpha\beta\delta} S_\Delta^{\beta\nu}(p'-z)
 (p'-z)_\mu\epsilon_\lambda
\gamma^{\mu\nu\lambda}(\slp'-\slk+m)\slq\gamma_5}{[z^2-m_\pi^2]p'\cdot k}\Bigg\}
\ket{p}.
\end{align}
The contribution of the diagrams corresponding to Fig.~\ref{fD1c} vanishes after dimensional regularization.
\begin{align}
\nn\mathcal{M}_{\ref{fD1d}} =&-\frac{eg_Ah_A^2}{12F_\pi^3M_{\Delta}^2}
\int{\frac{\mathrm{d}^dz}{(2\pi)^d}}
\bra{p',q}\Bigg\{\\
\nn &\frac{(p+k)_\alpha q_\delta\gamma^{\alpha\beta\delta} S_\Delta^{\beta\nu}(p+k)(p+k)_\mu\epsilon_\lambda
\gamma^{\mu\nu\lambda}
(\slp-\slz+m)\slz\gamma_5}{[z^2-m_\pi^2][(p-z)^2-m^2]}\\
&+\frac{\slz\gamma_5(\slp'-\slz+m) (p'-k)_\alpha\epsilon_\delta
\gamma^{\alpha\beta\delta}
 S_\Delta^{\beta\nu}(p'-k)(p'-k)_\mu q_\lambda\gamma^{\mu\nu\lambda}}{[z^2-m_\pi^2][(p'-z)^2-m^2]}
\Bigg\}\ket{p},\\[1em]
\nn \mathcal{M}_{\ref{fD1e}} =&\frac{eg_Ah_A^2}{12F_\pi^3M_{\Delta}^2}
\int{\frac{\mathrm{d}^dz}{(2\pi)^d}}
\bra{p',q}\Bigg\{\\
\nn&\frac{\epsilon\cdot z \slq\gamma_5(\slp+\slk+m)(p-z)_\alpha (z+k)_\delta\gamma^{\alpha\beta\delta} 
 S_\Delta^{\beta\nu}(p-z)(p-z)_\mu z_\lambda\gamma^{\mu\nu\lambda}}{[z^2-m_\pi^2][(z+k)^2-m_\pi^2]p\cdot k}\\
&-\frac{\epsilon\cdot z(p'-k-z)_\alpha (z+k)_\delta\gamma^{\alpha\beta\delta}  S_\Delta^{\beta\nu}(p'-k-z)
(p'-k-z)_\mu z_\lambda\gamma^{\mu\nu\lambda}(\slp'-\slk+m)\slq\gamma_5}{[z^2-m_\pi^2][(z+k)^2-m_\pi^2]p'\cdot k}\Bigg\}
\ket{p},\\[1em]
\nn \mathcal{M}_{\ref{fD1f}} =&\frac{eg_Ah_A^2}{6F_\pi^3M_{\Delta}^2}
\int{\frac{\mathrm{d}^dz}{(2\pi)^d}}
\bra{p',q}\Bigg\{\\
\nn &\frac{\epsilon\cdot z(p+k)_\alpha q_\delta\gamma^{\alpha\beta\delta}  S_\Delta^{\beta\nu}(p+k)
(p+k)_\mu (z+k)_\lambda\gamma^{\mu\nu\lambda}(\slp-\slz+m)\slz\gamma_5}{[z^2-m_\pi^2][(z+k)^2-m_\pi^2][(p-z)^2-m^2]}\\
&+\frac{\epsilon\cdot z(\slz+\slk)\gamma_5(\slp'-\slk-\slz+m)(p'-k)_\alpha z_\delta\gamma^{\alpha\beta\delta}
 S_\Delta^{\beta\nu}(p'-k)(p'-k)_\mu q_\lambda\gamma^{\mu\nu\lambda}}{[z^2-m_\pi^2][(z+k)^2-m_\pi^2][(p'-k-z)^2-m^2]}\Bigg\}
\ket{p},\\[1em]
\nn \mathcal{M}_{\ref{fD1g}} =&-\frac{eg_Ah_A^2}{12F_\pi^3M_{\Delta}^2}
\int{\frac{\mathrm{d}^dz}{(2\pi)^d}}
\bra{p',q}\Bigg\{\\
\nn &\frac{(p+k)_\alpha q_\delta\gamma^{\alpha\beta\delta}  S_\Delta^{\beta\nu}(p+k)
 (p+k)_\mu z_\lambda\gamma^{\mu\nu\lambda}(\slp+\slk-\slz+m)\sle(\slp-\slz+m)\slz\gamma_5}{[z^2-m_\pi^2][(p+k-z)^2-m^2][(p-z)^2-m^2]}\\
 &+\frac{\slz\gamma_5(\slp'-\slz+m)\sle(\slp'-\slk-\slz+m)(p'-k)_\alpha z_\delta\gamma^{\alpha\beta\delta}
   S_\Delta^{\beta\nu}(p'-k)(p'-k)_\mu q_\lambda\gamma^{\mu\nu\lambda}}{[z^2-m_\pi^2][(p'-z)^2-m^2][(p'-k-z)^2-m^2]} \Bigg\}
\ket{p},\\[1em]
\nn \mathcal{M}_{\ref{fD1h}} =&-\frac{eg_Ah_A^2}{24F_\pi^3M_{\Delta}^2}
\int{\frac{\mathrm{d}^dz}{(2\pi)^d}}
\bra{p',q}\Bigg\{\\
\nn &\frac{(p'-z)_\alpha z_\delta\gamma^{\alpha\beta\delta} S_\Delta^{\beta\nu}(p'-z)
(p'-z)_\mu q_\lambda\gamma^{\mu\nu\lambda}(\slp+\slk-\slz+m)\slz\gamma_5(\slp+\slk+m)\sle}{[z^2-m_\pi^2][(p+k-z)^2-m^2] p\cdot k}\\
\nn&+\frac{\slz\gamma_5(\slp'-\slz+m)(p+k-z)_\alpha q_\delta\gamma^{\alpha\beta\delta} S_\Delta^{\beta\nu}(p+k-z)
(p+k-z)_\mu z_\lambda\gamma^{\mu\nu\lambda}(\slp+\slk+m)\sle}{[z^2-m_\pi^2][(p'-z)^2-m^2] p\cdot k}\\
\nn&-\frac{\sle(\slp'-\slk+m)(p'-k-z)_\alpha z_\delta\gamma^{\alpha\beta\delta} S_\Delta^{\beta\nu}(p'-k-z)
(p'-k-z)_\mu q_\lambda\gamma^{\mu\nu\lambda}(\slp-\slz+m)\slz\gamma_5}{[z^2-m_\pi^2][(p-z)^2-m^2] p'\cdot k}\\
 &-\frac{\sle(\slp'-\slk+m)\slz\gamma_5(\slp'-\slk-\slz+m)(p-z)_\alpha q_\delta\gamma^{\alpha\beta\delta} 
S_\Delta^{\beta\nu}(p-z)(p-z)_\mu z_\lambda\gamma^{\mu\nu\lambda}}{[z^2-m_\pi^2][(p'-k-z)^2-m^2] p'\cdot k}\Bigg\}	
\ket{p},\\[1em]
\nn \mathcal{M}_{\ref{fD1i}} =&\frac{eg_Ah_A^2}{12F_\pi^3M_{\Delta}^2}
\int{\frac{\mathrm{d}^dz}{(2\pi)^d}}
\bra{p',q}\Bigg\{\\
\nn &\frac{(p'-z)_\alpha z_\delta\gamma^{\alpha\beta\delta} S_\Delta^{\beta\nu}(p'-z)
(p'-z)_\mu q_\lambda\gamma^{\mu\nu\lambda}(\slp+\slk-\slz+m)\sle\gamma_5}{[z^2-m_\pi^2][(p+k-z)^2-m^2]}\\
&+\frac{\sle\gamma_5(\slp'-\slk-\slz+m)(p-z)_\alpha q_\delta\gamma^{\alpha\beta\delta} 
S_\Delta^{\beta\nu}(p-z)(p-z)_\mu z_\lambda\gamma^{\mu\nu\lambda}}{[z^2-m_\pi^2][(p'-k-z)^2-m^2]}
\Bigg\}\ket{p},\\[1em]
\nn\mathcal{M}_{\ref{fD1j}} =&\frac{eg_Ah_A^2}{12F_\pi^3M_{\Delta}^2}
\int{\frac{\mathrm{d}^dz}{(2\pi)^d}}
\bra{p',q}\Bigg\{\\
\nn &\frac{\slz\gamma_5(\slp'-\slz+m)(p+k-z)_\alpha q_\delta\gamma^{\alpha\beta\delta} S_\Delta^{\beta\nu}(p+k-z)
(p+k-z)_\mu\epsilon_\lambda
\gamma^{\mu\nu\lambda}}{[z^2-m_\pi^2][(p'-z)^2-m^2]}\\
 &+\frac{(p'-k-z)_\alpha\epsilon_\delta
\gamma^{\alpha\beta\delta} S_\Delta^{\beta\nu}(p'-k-z)
(p'-k-z)_\mu q_\lambda\gamma^{\mu\nu\lambda}(\slp-\slz+m)\slz\gamma_5}{[z^2-m_\pi^2][(p-z)^2-m^2]}
\bigg\}\ket{p}.
\end{align}
The diagrams of Fig.~\ref{fD1k} do not contribute to the amplitude at the considered order, due to isospin cancellation.
\begin{align}
\nn \mathcal{M}_{\ref{fD1l}} =&-\frac{eg_Ah_A^2}{6F_\pi^3M_{\Delta}^2}
\int{\frac{\mathrm{d}^dz}{(2\pi)^d}}
\bra{p',q}\Bigg\{\\
\nn &\frac{\epsilon\cdot z(p'-k-z)_\alpha (z+k)_\delta\gamma^{\alpha\beta\delta} S_\Delta^{\beta\nu}(p'-k-z)
(p'-k-z)_\mu q_\lambda\gamma^{\mu\nu\lambda}
 (\slp-\slz+m)\slz\gamma_5}{[z^2-m_\pi^2][(z+k)^2-m_\pi^2][(p-z)^2-m^2]}\\
\nn &+\frac{\epsilon\cdot z(\slz+\slk)\gamma_5(\slp'-\slk-\slz+m)(p-z)_\alpha q_\delta\gamma^{\alpha\beta\delta}
S_\Delta^{\beta\nu}(p-z)(p-z)_\mu z_\lambda\gamma^{\mu\nu\lambda}}{[z^2-m_\pi^2][(z+k)^2-m_\pi^2][(p'-k-z)^2-m^2]}
\Bigg\}\ket{p},\\[1em]
\nn \mathcal{M}_{\ref{fD1m}} =&-\frac{eg_Ah_A^2}{12F_\pi^3M_{\Delta}^2}
\int{\frac{\mathrm{d}^dz}{(2\pi)^d}}
\bra{p',q}\Bigg\{\\
\nn &\frac{(p'-z)_\alpha z_\delta\gamma^{\alpha\beta\delta} S_\Delta^{\beta\nu}(p'-z)
(p'-z)_\mu q_\lambda\gamma^{\mu\nu\lambda}
 (\slp+\slk-\slz+m)\sle(\slp-\slz+m)\slz\gamma_5}{[z^2-m_\pi^2][(p+k-z)^2-m^2][(p-z)^2-m^2]}\\
 &+\frac{\slz\gamma_5(\slp'-\slz+m)\sle(\slp-\slk-\slz+m)(p-z)_\mu q_\lambda\gamma^{\mu\nu\lambda}
 S_\Delta^{\beta\nu}(p-z)(p-z)_\alpha z_\delta\gamma^{\alpha\beta\delta}}{[z^2-m_\pi^2][(p'-z)^2-m^2][(p'-k-z)^2-m^2]}
\Bigg\}\ket{p}.
\end{align}


\begin{thebibliography}{99}
\section*{Bibliography}


\bibitem{Kroll} 
N. M. Kroll and M. A. Ruderman, Phys. Rev. \  {\bf 93}, 233 (1954).

\bibitem{DeBaenst:1971hp}
  P.~De Baenst,
  Nucl.\ Phys.\ B {\bf 24} (1970) 633.

\bibitem{Vainshtein:1972ih}
  A.~I.~Vainshtein and V.~I.~Zakharov,
  Nucl.\ Phys.\ B {\bf 36} (1972) 589.

\bibitem{Mazzucato:1986dz} 
  E.~Mazzucato {\it et al.},
  Phys.\ Rev.\ Lett.\  {\bf 57} (1986) 3144.

\bibitem{Beck:1990da}
  R.~Beck {\it et al.},
  Phys.\ Rev.\ Lett.\  {\bf 65} (1990) 1841.


\bibitem{Drechsel:1992pn} 
  D.~Drechsel and L.~Tiator,
  J.\ Phys.\ G {\bf 18} (1992) 449.


\bibitem{Bernard:2006gx} 
  V.~Bernard and U.~-G.~Meissner,
  Ann.\ Rev.\ Nucl.\ Part.\ Sci.\  {\bf 57} (2007) 33.


\bibitem{Bernard:1991rt}
  V.~Bernard, N.~Kaiser, J.~Gasser and U.~G.~Meissner,
  Phys.\ Lett.\ B {\bf 268} (1991) 291.

\bibitem{Bernard:1992nc}
  V.~Bernard, N.~Kaiser and U.~G.~Meissner,
  Nucl.\ Phys.\ B {\bf 383} (1992) 442.

\bibitem{Jenkins:1990jv}
  E.~E.~Jenkins and A.~V.~Manohar,
Lagrangian,''
  Phys.\ Lett.\ B {\bf 255} (1991) 558.

\bibitem{Jenkins:1991es}
  E.~E.~Jenkins and A.~V.~Manohar,
  Phys.\ Lett.\ B {\bf 259} (1991) 353.

\bibitem{Gasser:1987rb}
  J.~Gasser, M.~E.~Sainio and A.~Svarc,
  Nucl.\ Phys.\ B {\bf 307} (1988) 779.

\bibitem{Bernard:2001gz} 
  V.~Bernard, N.~Kaiser and U.~-G.~Meissner,
  Eur.\ Phys.\ J.\ A {\bf 11} (2001) 209.

\bibitem{Hornidge:2012ca} 
  D.~Hornidge {\it et al.}  [A2 and CB-TAPS Collaborations],
  Phys.\ Rev.\ Lett.\  {\bf 111}, no. 6 (2013) 062004.


%
\bibitem{FernandezRamirez:2012nw} 
  C.~Fernandez-Ramirez and A.~M.~Bernstein,
  Phys.\ Lett.\ B {\bf 724} (2013) 253.


\bibitem{Gegelia:1999gf}
  J.~Gegelia and G.~Japaridze,
  Phys.\ Rev.\ D {\bf 60} (1999) 114038.

\bibitem{Fuchs:2003qc}
  T.~Fuchs, J.~Gegelia, G.~Japaridze and S.~Scherer,
  Phys.\ Rev.\ D {\bf 68} (2003) 056005.

  

\bibitem{Fuchs:2003ir}
  T.~Fuchs, J.~Gegelia and S.~Scherer,
  J.\ Phys.\ G {\bf 30} (2004) 1407.

\bibitem{Lehnhart:2004vi}
  B.~C.~Lehnhart, J.~Gegelia and S.~Scherer,
  J.\ Phys.\ G {\bf 31} (2005) 89.

\bibitem{Schindler:2006it}
  M.~R.~Schindler, T.~Fuchs, J.~Gegelia and S.~Scherer,
  Phys.\ Rev.\ C {\bf 75} (2007) 025202.

\bibitem{Schindler:2006ha}
  M.~R.~Schindler, D.~Djukanovic, J.~Gegelia and S.~Scherer,
  Phys.\ Lett.\ B {\bf 649} (2007) 390.


\bibitem{Geng:2008mf}
  L.~S.~Geng, J.~Martin Camalich, L.~Alvarez-Ruso and M.~J.~Vicente Vacas,
  Phys.\ Rev.\ Lett.\  {\bf 101} (2008) 222002.

\bibitem{Geng:2009ik}
  L.~S.~Geng, J.~Martin Camalich and M.~J.~Vicente Vacas,
  Phys.\ Rev.\ D {\bf 79} (2009) 094022.

\bibitem{MartinCamalich:2010fp}
  J.~Martin Camalich, L.~S.~Geng and M.~J.~Vicente Vacas,
  Phys.\ Rev.\ D {\bf 82} (2010) 074504.

\bibitem{Alarcon:2011zs}
  J.~M.~Alarcon, J.~Martin Camalich and J.~A.~Oller,
  Phys.\ Rev.\ D {\bf 85} (2012) 051503.

\bibitem{Ledwig:2012}
  T.~Ledwig, J.~M.~Camalich, V.~Pascalutsa and M.~Vanderhaeghen,
  Phys.\ Rev.\ D {\bf 85} (2012) 034013.

\bibitem{Chen:2012nx}
  Y.~H.~Chen, D.~L.~Yao and H.~Q.~Zheng,
  Phys.\ Rev.\ D {\bf 87}, no. 5 (2013)  054019.

\bibitem{Ledwig:2013}
L.~Alvarez-Ruso, T.~Ledwig, J.~M.~Camalich and M.~J.~Vicente-Vacas,
 Phys.\ Rev.\ D {\bf 88}, no. 5 (2013) 054507.

\bibitem{Ledwig:2014rfa}
  T.~Ledwig, J.~M.~Camalich, L.~S.~Geng and M.~J.~V.~Vacas,
  Phys.\ Rev.\ D {\bf 90} (2014) 054502.

\bibitem{Lensky:2014dda}
      V.~Lensky, J.~M.~Alarc\'on and V.~Pascalutsa,
      Phys.\ Rev.\ C {\bf 90}, no. 5 (2014) 055202.

\bibitem{Hilt:2013uf} 
  M.~Hilt, S.~Scherer and L.~Tiator,
  Phys.\ Rev.\ C {\bf 87}, no. 4 (2013) 045204.

\bibitem{Ericson:1988gk}
  T.~E.~O.~Ericson and W.~Weise,
  OXFORD, UK: CLARENDON (1988) 479 P (THE INTERNATIONAL SERIES OF MONOGRAPHS ON PHYSICS, 74).


\bibitem{Hemmert:1996xg} 
  T.~R.~Hemmert, B.~R.~Holstein and J.~Kambor,
  Phys.\ Lett.\ B {\bf 395} (1997) 89.


\bibitem{Pascalutsa:2004pk} 
  V.~Pascalutsa and J.~A.~Tjon,
  Phys.\ Rev.\ C {\bf 70} (2004) 035209.

\bibitem{Pascalutsa:2005vq} 
  V.~Pascalutsa and M.~Vanderhaeghen,
  Phys.\ Rev.\ D {\bf 73} (2006) 034003.

\bibitem{FernandezRamirez:2005iv}
  C.~Fernandez-Ramirez, E.~Moya de Guerra and J.~M.~Udias,
  Annals Phys.\  {\bf 321} (2006) 1408.

\bibitem{Lensky:2009uv}
  V.~Lensky and V.~Pascalutsa,
  Eur.\ Phys.\ J.\ C {\bf 65} (2010) 195.
%

\bibitem{Blin:2015era} 
  A.~H.~Blin, T.~Gutsche, T.~Ledwig and V.~E.~Lyubovitskij,
  Phys.\ Rev.\ D {\bf 92}, no. 9 (2015) 096004.

\bibitem{Alarcon:2012kn}
  J.~M.~Alarcon, J.~Martin Camalich and J.~A.~Oller,
  Annals Phys.\  {\bf 336} (2013) 413.

\bibitem{Blin:2015} 
 A.~N.~Hiller Blin, T.~Ledwig and M.~J.~Vicente Vacas,
 Phys.\ Lett.\ B {\bf 747} (2015) 217.

\bibitem{Cawthorne:2015} 
  L.~W.~Cawthorne and J.~A.~McGovern,
  arXiv:1510.09136.

\bibitem{Chew:1957tf}
  G.~F.~Chew, M.~L.~Goldberger, F.~E.~Low and Y.~Nambu,
  Phys.\ Rev.\  {\bf 106} (1957) 1345.

\bibitem{Fettes:2000gb} 
  N.~Fettes, U.~G.~Meissner, M.~Mojzis and S.~Steininger,
  Annals Phys.\  {\bf 283} (2000) 273
  [Erratum-ibid.\  {\bf 288} (2001) 249].

\bibitem{Weinberg:1978kz}
  S.~Weinberg,
  Physica A {\bf 96} (1979) 327.

\bibitem{Gasser:1983yg}
  J.~Gasser and H.~Leutwyler,
  Annals Phys.\  {\bf 158} (1984) 142.

\bibitem{Gasser:1984gg}
  J.~Gasser and H.~Leutwyler,
  Nucl.\ Phys.\ B {\bf 250} (1985) 465.






\bibitem{Pascalutsa:1998pw}
  V.~Pascalutsa,
  Phys.\ Rev.\ D {\bf 58} (1998) 096002.

\bibitem{Pascalutsa:1999zz}
  V.~Pascalutsa and R.~Timmermans,
  Phys.\ Rev.\ C {\bf 60} (1999) 042201.

\bibitem{Pascalutsa:2000kd}
  V.~Pascalutsa,
  Phys.\ Lett.\ B {\bf 503} (2001) 85.

\bibitem{Pascalutsa:2006up}
  V.~Pascalutsa, M.~Vanderhaeghen and S.~N.~Yang,
  Phys.\ Rept.\  {\bf 437} (2007) 125.

\bibitem{FORM:2000} 
 J.~A.~M.~Vermaseren,
  math-ph/0010025.

\bibitem{FORM:2012} 
J.~Kuipers, T.~Ueda, J.~A.~M.~Vermaseren and J.~Vollinga,
 Comput.\ Phys.\ Commun. {\bf184} (2013) 1453.

\bibitem{FC:1991} 
R.~Mertig, M.~B\"ohm and A.~Denner,
 Comput.\ Phys.\ Commun. {\bf64} (1991) 345.

\bibitem{FC:2016} 
 V.~Shtabovenko, R.~Mertig and F.~Orellana,
 arXiv:1601.01167.

\bibitem{Scherer:2012xha}
  S.~Scherer and M.~R.~Schindler,
  Lect.\ Notes Phys.\  {\bf 830} (2012) 1.

\bibitem{Hilt:2013fda} 
  M.~Hilt, B.~C.~Lehnhart, S.~Scherer and L.~Tiator,
  Phys.\ Rev.\ C {\bf 88}, (2013) 055207.


\bibitem{Pascalutsa:2002pi} 
  V.~Pascalutsa and D.~R.~Phillips,
  Phys.\ Rev.\ C {\bf 67} (2003) 055202.



\bibitem{Pascalutsa:2005ts} 
  V.~Pascalutsa and M.~Vanderhaeghen,
  Phys.\ Rev.\ Lett.\  {\bf 95} (2005) 232001.

\bibitem{Hornidge:pri} 
D.~Hornidge, private communication.

\bibitem{FernandezRamirez:2009su} 
  C.~Fernandez-Ramirez, A.~M.~Bernstein and T.~W.~Donnelly,
  Phys.\ Lett.\ B {\bf 679}, (2009) 41.


\bibitem{FernandezRamirez:2009jb} 
  C.~Fernandez-Ramirez, A.~M.~Bernstein and T.~W.~Donnelly,
  Phys.\ Rev.\ C {\bf 80}, (2009) 065201.




\end{thebibliography}
\end{document}